\documentclass{aa}  

\usepackage{graphicx}
\usepackage{siunitx}
\usepackage{float}
\usepackage{xcolor}
\usepackage{amsmath}
\usepackage{subcaption}
\usepackage{multirow}
\usepackage{txfonts}
\begin{document} 

   \title{On the distribution of the the near-solar bound dust grains detected with Parker Solar Probe}

      \author{S. Ko\v{c}i\v{s}\v{c}\'{a}k
          \inst{1}\fnmsep\thanks{samuel.kociscak@uit.no}
          \and
          A. Theodorsen
          \inst{1}
          \and
          I. Mann
          \inst{1}
          }

   \institute{
   Department of Physics and Technology, UiT The Arctic University of Norway, 9037, Troms{\o}, Norway
         }

   \date{Received Month DD, YYYY; accepted Month DD, YYYY}
 
  \abstract
   {Parker Solar Probe (PSP) counts dust impacts in the near-solar region, but modelling effort is needed to understand the dust population's properties.}
   {We aim to constrain the dust cloud's properties based on the flux observed by PSP.}
   {We develop a forward-model for the bound dust detection rates using the formalism of 6D phase space distribution of the dust. We apply the model to the location table of different PSP's solar encounter groups. We explain some of the near-perihelion features observed in the data as well as the broader characteristic of the dust flux between $\SI{0.15}{AU}$ and $\SI{0.5}{AU}$. We compare the measurements of PSP to the measurements of Solar Orbiter (SolO) near $\SI{1}{AU}$ to expose the differences between the two spacecraft.}
   {We found that the dust flux observed by PSP between $\SI{0.15}{AU}$ and $\SI{0.5}{AU}$ in post-perihelia can be explained by dust on bound orbits and is consistent with a broad range of orbital parameters, including dust on circular orbits. However, the dust number density as a function of the heliocentric distance and the scaling of detection efficiency with the relative speed are important to explain the observed flux variation. The data suggest that the slope of differential mass distribution $\delta$ is between $0.14$ and $0.49$. The near-perihelion observations, however, show the flux maxima, which are inconsistent with the circular dust model, and additional effects may play a role. We found indication that the sunward side of PSP is less sensitive to the dust impacts, compared to the other PSP's surfaces.}
   {We show that the dust flux on PSP can be explained by non-circular bound dust and the detection capabilities of PSP. The scaling of flux with the impact speed is especially important, and shallower than previously assumed.}


   \maketitle

\section{Introduction}

The Parker Solar Probe (\textit{PSP}) \citep{fox2016solar} and Solar Orbiter (\textit{SolO}) \citep{muller2020solar} space missions are currently exploring the inner Solar System, conducting in-situ measurements on an unprecedented scale and traversing regions never before reached by space probes. Through remote and in-situ detections, they among others also enable unprecedented observations of the innermost region of the interplanetary dust cloud. Most of the interplanetary dust cloud originates from the asteroids and the comets of the solar system and it is observed in the Zodiacal light and its inner extension into the solar corona, called F-corona (see \citet{koschny2019interplanetary} for a recent review). While there were only a few in-situ measurements of dust within $\SI{1}{AU}$, this changed with SolO and PSP because their experiments investigating plasma waves with antenna measurements: FIELDS \citep{bale2016fields} on PSP and RPW \citep{maksimovic2020solar} on SolO are also sensitive to dust impacts. Earlier analyses of impact measurements of PSP \citep{szalay2020near,szalay2021collisional,malaspina2020situ} and SolO \citep{zaslavsky2012interplanetary,kociscak2023modeling} showed that the observation included mainly dust in hyperbolic trajectories, which were carried away from the Sun due to the effect of the radiation pressure force, and dust in bound orbits determined by gravity. The hyperbolic grains, which are pushed outward as a result of the radiation pressure force are often denoted as $\beta$-meteoroids, since for those particles the ratio $\beta$ of the radiation pressure force $F_{rp}$ and the gravity force $F_g$ is roughly $0.5$ or larger. Their motion is the typical central force problem, while the effective gravity force the grains are subjected to is reduced by a factor of $(1-\beta)$. Electromagnetic forces seem to have little influence in comparison, or affect only a small fraction of the dust \citep{mann2021dust}. It is assumed that the observed dust is created by dust-dust collisions near the Sun. This would imply that the relative amount of $\beta$-meteoroids compared to dust in bound orbits decreases with increasing proximity to the Sun. We therefore investigate whether the observed dust fluxes in the close vicinity of the Sun can be due to dust in bound orbits. We also compare the dust fluxes that are observed with SolO and PSP.  

The work is structured as follows: in Sec.~\ref{ch:psp_and_solo}, we present Parker Solar Probe and Solar Orbiter and the previous results of other authors. We present the principle and the limitations of dust detection with electrical antennas in Sec.~\ref{ch:measurements} along with the the data from the two spacecraft, which we use in later analysis. In Sec.~\ref{ch:psp_vs_solo}, we compare the dust measurements of the two spacecraft near $\SI{1}{AU}$. In Sec.~\ref{ch:model} we introduce the parametric forward model, which we use to explain the features observed in the dust flux measured by Parker Solar Probe in Sec.~\ref{ch:results}. We discuss the implications of the results in Sec.~\ref{ch:discussion} and in Sec.~\ref{ch:conclusion} we summarize.

\section{Parker Solar Probe and Solar Orbiter} \label{ch:psp_and_solo}

\subsection{Dust detection with Parker Solar Probe}

PSP orbits the Sun on highly eccentric orbits between $\SI{0.05}{AU}$ and $\SI{1}{AU}$. The orbital parameters change significantly during gravity assists, but remain nearly identical between the assists, forming several distinct orbital groups of nearly identical orbits. PSP has made $20$ orbits around the Sun so far. PSP was found to be sensitive to impacts of dust grains on the spacecraft body \citep{szalay2020near,malaspina2020situ,page2020examining} thought the measurements of its FIELDS antenna suite \citep{bale2016fields}. In addition to the electrical antennas, dust phenomena were observed with the Wide-Field Imager for Parker Solar Probe (\textit{WISPR}) \citep{stenborg2021psp,malaspina2022clouds} and dust possibly damaged the Integrated Science Investigation of the Sun (\textit{IS}$\odot$\textit{IS}) instrument \citep{szalay2020near}.

It was found that the dust detection can be successfully modelled as a combination of bound dust and $\beta$-meteoroids \citep{szalay2021collisional}. During the initial orbits, most impacts were attributed to $\beta$-meteoroids. However, during the later orbits, in post-perihelia, where the relative speed between $\beta$-meteoroids and the PSP is low, the dust counts are sometimes likely to be dominated by bound dust. The $\beta$ value depends greatly on the grain's size and $\beta$-meteoroids have the typical size of $\SI{100}{nm} \lesssim d \lesssim \SI{1}{\mu m}$, since this is where $\beta$ has its maximum \citep{kimura2003elemental}. The Lorentz force is negligible for $\beta$-meteoroids and boud dust, due to the low $Q/m < \SI{1e-7}{e/m_p}$ \citep{czechowski2010formation}.

The size of the grains detected with antennas is estimated only indirectly. Based on the impact charge yields measured in laboratory hypervelocity experiments, and in combination with estimates of dust impact speed for individual dust populations based on first principles modelling, \citet{szalay2021collisional} estimated the lower size radius limit for detections as a function of time. They found that during the orbits $8 - 16$ of PSP, the bound dust grains were detected as small as $r \gtrsim \SI{300}{nm}$ near perihelia and $r \gtrsim \SI{2}{\mu m}$ near aphelia, while the $\beta$-meteoroids were detected as small as $r \gtrsim \SI{100}{nm}$ during most of the orbit, except for post-perihelia, where the threshold was close to $r \gtrsim \SI{1}{\mu m}$. We note that while $r = \SI{100}{nm}$ is close to the lower size limit of $\beta$-meteoroids, $r = \SI{1}{\mu m}$ is close to the upper size limit.

\cite{szalay2021collisional} observed a clear double peak structure with a minimum in perihelion during the orbits 4, 5, and 6 of PSP. The minimum was anticipated and at least partially explained by \cite{szalay2020near} as being due to alignment between the nearly-circular speed of bound dust and purely azimuthal speed of PSP in the perihelion. \cite{szalay2021collisional} also explained the post-perihelion maximum as being potentially due to the encounter between PSP and the hypothesized Geminids $\beta$-stream produced by the collisions between the bound dust cloud and the Gemminids meteoroid stream.

\subsection{Dust detection with Solar Orbiter} \label{ch:dust_solo}

PSP's observations of the inner zodiacal cloud are unique, and the closest available comparable observations are those of Solar Orbiter \citep{mann2019dust}. Similarly to PSP, SolO is equipped with electrical antennas of its Radio and Plasma Waves (\textit{RPW}) instrument \citep{maksimovic2020solar}, which registers dust impact on the body of the spacecraft through their electrical signatures \citep{soucek2021solar}. Compared to PSP, Solo experiences much lower radial speed, and with its perihelia of about $\SI{0.3}{AU}$, it doesn't go nearly as close to the Sun. Unlike PSP, the dust flux SolO measures is always dominated by $\beta$-meteoroids \citep{zaslavsky2021first,kociscak2023modeling}. Those were concluded to have the radius of $r \gtrsim \SI{100}{nm}$ \citep{zaslavsky2021first} and $\beta \gtrsim 0.5$ \citep{kociscak2023modeling}. 

\subsection{Components of the observed dust flux} \label{ch:populations_compared}

One of the difficulties in explaining the observed flux is that several populations contribute to the detections \citep{mann2019dust,szalay2020near,szalay2021collisional,kociscak2023modeling}, and therefore one must assume several components of the impact rate, which greatly decreases the fidelity of parameter estimation. As was debated by \cite{szalay2020near}, bound dust and $\beta$-meteoroids are the main contributors to the dust flux on PSP. Interstellar dust grains \citep{mann2010interstellar} were not yet reported by either of the spacecraft, but their presence in data is likely, even if they are a minor contributor to the overall detection counts.

A model for the $\beta$-meteoroid flux observed by both spacecraft must be currently built on many assumptions, as there are many unknowns to the population of $\beta$-meteoroids. Any differences between the fluxes might be attributed to the properties of the population, or to the differences between the two spacecraft, which are numerous. The $\beta$-meteoroids population was studied extensively by these \citep{szalay2021collisional,zaslavsky2021first,kociscak2023modeling} and by other spacecraft \citep{zaslavsky2012interplanetary,malaspina2014interplanetary}. Although we focus on the bound dust component, constraining this will implicitly provide information on $\beta$-meteoroids, since they together make up the detected flux. 

Bound dust grains are in bound orbits, and therefore have both positive and negative heliocentric speeds. In the special case of a circular orbit, the dust grain has zero heliocentric component of speed. $\beta$-meteoroids are on outbound trajectories, with each of them having a positive heliocentric speed. The proportion of $\beta$-meteoroids is the highest, when the spacecraft has negative heliocentric speed. Conversely, the proportion of impacts of bound dust to all impacts is the highest when PSP's radial speed is positive. The relative speed between PSP and bound dust is approximated well by PSP's radial speed \citep{szalay2020near}, which was between $\SI{32}{km s^{-1}}$ and $\SI{72}{km s^{-1}}$ during the orbital groups $1-5$ between $\SI{0.15}{AU}$ and $\SI{0.5}{AU}$. The relative speed between PSP and $\beta$-meteoroids depends on their outward speed and the creation region, and was in tens of $\si{km/s}$ in the outbound legs of the studied orbits.

\citeauthor{szalay2020near} assumed perfectly circular bound dust trajectories with $\beta = 0$ and $\beta$-meteoroids originating at $R_0 = 5R_{sun}$ and having $\beta=0.5$. Under these assumptions, they found the relative speed between PSP and bound dust to be higher at $\SI{0.15}{AU} < R < \SI{0.5}{AU}$ for the 6th and subsequent orbits, compared to $\beta$-meteoroids. A two-component fit to the data performed by \citep{szalay2021collisional} is consistent with the flux of bound dust being higher than the flux of $\beta$ for $\SI{0.15}{AU} < R < \SI{0.5}{AU}$ during the 6th orbit. In fact, the fit suggests that bound dust flux is more than a decade higher than in $\beta$-meteoroid flux for the 6th perihelion's outbound leg at $R=\SI{0.2}{AU}$. Although the exact numbers and distances are model specific, the general trend is clear: the bound dust flux is higher than $\beta$-meteoroid flux for a good portion of the post-perihelion passage, especially for the orbit six and the later orbits.

\section{Measurement technique} \label{ch:measurements}

\subsection{Antenna dust detection} \label{ch:principle}

When a dust grain collides with a spacecraft at a speed, which exceeds a few $\si{km/s}$, the impact is followed by a release of a plume of quasi-neutral charge cloud \citep{friichtenicht1962two}. The amount of this charge depends on many factors, most importantly the grain's material and mass, and the impact speed. Depending on the spacecraft and the surrounding environment, a portion of the charge is collected by the spacecraft and the rest escapes from its vicinity. This process, which usually happens in $\si{\mu s}$, may be detected with fast measurements of electrical antennas, if such measurement is present. In this way, electrical antennas performing fast measurements act as dust detectors, while the whole surface of the spacecraft's body is potentially sensitive to the impacts. 

The configuration of the antennas, their location with respect to the impact site, and the material of the spacecraft surface are among the factors, which influence the detection efficiency the most \citep{shen2023variability,collette2014micrometeoroid}. Both the FIELDS instrument of PSP and the RPW instrument SolO are equipped with multiple thick cylindrical antennas close to their respective sun-facing heat shields \citep{bale2016fields,maksimovic2020solar}. Some of these antennas operate in the monopole configuration for both spacecraft, in which the voltage is measured between the antenna and the spacecraft body. This is the preferable configuration for dust detection, since it makes the body a more sensitive target \citep{meyer2014importance}, in comparison to dipole antennas. In dipole configuration, voltage is measured between two antennas, and such measurement is not directly sensitive to the potential of the body.

Since detections are recorded on the body of the spacecraft, the effective cross section of the body is important to establish. One of the differences between PSP and SolO is the shape. The body of SolO, excluding the solar panels, has a rough cuboid shape \citep{solo_model}. The body of PSP, excluding the solar panels, has more cylindrical shape \citep{psp_model}. The difference between a cuboid and a cylinder is of no consequence to dust modelling within the plane of ecliptics, but is a factor if there is an inclination between the spacecraft's trajectory and that of the dust.

\subsection{PSP's potential} \label{sec:potential}

\begin{figure}[h]
 	\centering
 	\includegraphics[width=9cm]{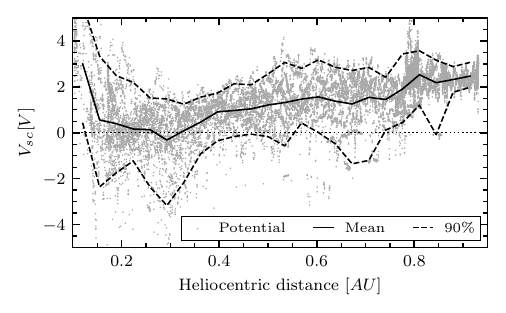}
 	\caption{Spacecraft potential estimated as the average of four $DC$ antenna measurements.}
 	\label{fig:potentials}
\end{figure}

The potential of the spacecraft has influence on the amplitude of the generated signal, which was studied in laboratory previously \citep{collette2016characteristic,shen2023variability}, and this affects the detection efficiency. In the absence of a direct measurement, the floating potential of PSP can be approximated by the average of the DC voltages $V_i$ between the antenna $i$ and the spacecraft \citep{bale2020dust}. If the antennas are on the local plasma potential, then the antennas measure voltage between the spacecraft body and the ambient plasma. The dependence of the spacecraft potential on the heliocentric distance is shown in Fig.~\ref{fig:potentials}. To reduce the amount of data to show, points were drawn uniformly randomly from the first $30$ months of the mission and the potential was inferred this way, using \verb|DFB_WF_DC| data product of FIELDS \citep{malaspina2016digital}. There are many factors beyond the heliocentric distance, which influence the final potential \citep{guillemant2012solar,guillemant2013simulation}. Even still, one can see that the potential is mostly positive outside of $\SI{0.3}{AU}$, close to zero at around $\SI{0.2}{AU}$, and changing suddenly inward of $\SI{0.15}{AU}$. The potential of the TPS heat shield is presumably different again, since its sunward side is not conductively coupled to the spacecraft, as is discussed in Sec.~\ref{ch:heat_shield}. Nevertheless, this data suggests that the dust detection process does not change significantly with the heliocentric distance, if the spacecraft is outside of $\SI{0.15}{AU}$. This distance coincides with the distance of $\approx \SI{0.16}{AU}$, inside which the heat shield was estimated to become conductively coupled to the spacecraft body \citep{diaz2021parker}. These two distances are possibly related. 

\subsection{Influence of the heat shield} \label{ch:heat_shield}

\begin{figure}[h]
 	\centering
 	\includegraphics[width=9cm]{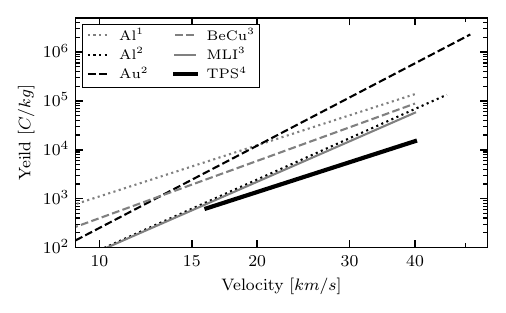}
 	\caption{Mass-normalized impact charge yield for several common spacecraft materials, assuming $10^{-14} \, \si{kg}$ dust. TPS stands for PSP's Thermal Protection System and MLI stands for the multilayer insulation of Solar Terretrial Relations Observatory (STEREO). References: 1 -- \cite{mcbride1999meteoroid}, 2 -- \cite{grun1984impact}, 3 -- \cite{collette2014micrometeoroid}, 4 -- \cite{shen2021phd}. All experiments were done with iron grains.}
 	\label{fig:charge_yield}
\end{figure}

Both PSP and SolO are protected against the extreme near-solar conditions by heat shields. The heat shields have different surface materials than the rest of the spacecraft. The sunward side of PSP's heat shield (Thermal Protection System --- \textit{TPS}) is, in addition, on a different potential to the rest of the body. These both influence dust detection. The sunward side of PSP's heat shield is made of alumina, which is non-conductive nor is it conductively connected to the spacecraft body \citep{reynolds2013solar,diaz2021parker}. The shield only becomes conductively coupled to the spacecraft body through plasma currents, once the spacecraft is inside of $\approx \SI{0.16}{AU}$ \citep{diaz2021parker}. The sunward side of SolO's heat shield is made of titanium and is conductively coupled to the rest of the spacecraft body \citep{damasio2015thermal}. The heat shields are exposed to dust impacts and even impacts on the non-conductive heat shield of PSP generate impact plasma \citep{shen2021phd}, which is potentially identified in the antenna measurements. Unlike impacts in the spacecraft parts connected conductively with the body ground, impacts on the heat shield only produce dipole response, which is more directionally dependent \citep{shen2023variability} and generally weaker \citep{mann2019dust}. Moreover, the amount of charge generated by impacts on the PSP's heat shield is comparably lower, with respect to other common spacecraft materials, see Fig.~\ref{fig:charge_yield}. 

\subsection{Data}

In the present work, we use the data of FIELDS-detected dust counts along the trajectory of PSP, made available by \cite{malaspina2023dust}. The data are built on the \verb|TDSmax| data product of the FIELDS instrument \citep{bale2016fields}, and assume that all the fast electrical phenomena strong enough in the monopole measurement detected over quiet enough periods of time contain dust impacts. This was demonstrated to be a good approximation of the actual dust count, with about $\SI{10.5}{\%}$ false positive rate and about $\SI{7}{\%}$ miss rate \citep{malaspina2023dust}. The data are structured in intervals of $\SI{8}{h}$, and contain the impact count in the time interval corrected for under-counting. The data also contain the total effective observation time in the $\SI{8}{h}$ interval, corrected for the time then wave activity made dust detection ineffective. The data product is described in detail by \cite{malaspina2023dust}. The data from the orbits $1-16$, and, therefore from the first five orbital groups are examined. 

In addition to the PSP data, we also use SolO dust data. SolO detects dust with the electrical antennas of RPW instrument, and we use the convolutional neural network (\textit{CNN}) data product provided by \citep{kvammen2022github}. The data product builds on the data set of time-domain sampled triggered electrical waveform data and was shown to have a low false positive error rate of about $\SI{4}{\%}$ and about $\SI{3}{\%}$ of miss rate \citep{kvammen2022convolutional}.

\section{PSP and SolO dust flux comparison} \label{ch:psp_vs_solo}

Although SolO and PSP have very different orbits at any given time, a direct comparison of dust fluxes can be done for several points near $\SI{1}{AU}$, where the two spacecraft had a similar heliocontric distance and speed --- albeit at different times and different helio-ecliptic latitudes. Six such time intervals were found and are listed in Tab.~\ref{tab:case_study}. Three of these are in pre-perihelion, with negative heliocentric radial speed ($v_r<0$) and three are in post-perihelion ($v_r>0$). In pre-perihelion, the ram direction of the spacecraft lies between azimuthal and sunward. The proportion of impacts on the heat shield is likely higher, than in post-perihelion, when the ram-direction lies between azimuthal and anti-sunward. 

\begin{figure}[h]
 	\centering
 	\includegraphics[width=9cm]{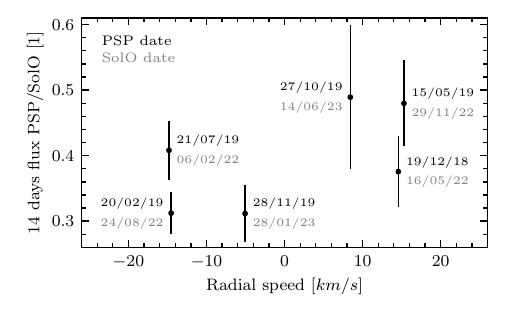}
 	\caption{Comparison of the 14 days cumulative flux (centered at the indicated day) observed with PSP and with SolO during their near alignments. The error bars are $\pm \sigma$, bootstrapped assuming Poisson distribution. See Tab.~\ref{tab:case_study} for the heliocentric locations, velocity components and fluxes corresponding to the individual points.}
 	\label{fig:case_study}
\end{figure}

We compare the $14$-day average flux $F$ during the six alignments ($7$ days before and $7$ days after the alignment), and the comparison is shown in Fig.~\ref{fig:case_study}. The flux implied by SolO is by a factor of two to three higher than the flux implied by PSP. Since the flux per unit area and time are compared, the ratio depends on the cross sections, which are here assumed $\SI{6.11}{m^2}$ and $\SI{10.34}{m^2}$ for PSP and SolO respectively. A part of the difference might be due to the difference between instruments and detection algorithms, leading to a different size sensitivity. The detection algorithm on PSP's FIELDS for example applies a signal threshold of $\SI{50}{mV}$ \citep{malaspina2023dust}, which has to be surpassed, in order for detection to count. This threshold, in combination with the dust speed and mass distribution, influences the total detected counts. 

One can see from Fig.~\ref{fig:case_study} that the relative detection rate of PSP with respect to SolO is higher in post-perihelia than in pre-perihelia. Unlike PSP, SolO's body is covered with conductive materials on all sides. If SolO is assumed to be equally sensitive to dust impacts from all sides, this implies that the sunward side of PSP is less sensitive than the rest of the spacecraft. This is possibly related to the non-conductive nature of the PSP's heat shield. A less sensitive heat-shield would also contribute to the lower overall flux through reducing the effective cross section of PSP. 

\begin{table*}[t]
\caption{Near alignments of speed and heliocentric distance between PSP and SolO.}
\centering
\label{tab:case_study}
\begin{tabular}{cccccc|cccccc}
\hline\hline
\multicolumn{1}{p{1.44cm}}{ \centering  \multirow{2}{*}{ PSP date } } & 
\multicolumn{1}{p{0.75cm}}{ \centering $R$ } & 
\multicolumn{1}{p{0.5cm}}{ \centering $\lambda$ } & 
\multicolumn{1}{p{1.18cm}}{ \centering $v_{r}$} & 
\multicolumn{1}{p{1.18cm}}{ \centering $v_{\phi}$} & 
\multicolumn{1}{p{1.35cm}}{ \centering $F$} \vline & 
\multicolumn{1}{p{1.57cm}}{ \centering  \multirow{2}{*}{ SolO date } } & 
\multicolumn{1}{p{0.75cm}}{ \centering $R$ } & 
\multicolumn{1}{p{0.5cm}}{ \centering $\lambda$ } & 
\multicolumn{1}{p{1.18cm}}{ \centering $v_{r}$} & 
\multicolumn{1}{p{1.18cm}}{ \centering $v_{\phi}$} & 
\multicolumn{1}{p{1.35cm}}{ \centering $F$} \\
& 
\multicolumn{1}{p{0.75cm}}{ \centering $[\si{AU}]$ } & 
\multicolumn{1}{p{0.5cm}}{ \centering $[\si{^\circ}]$ } & 
\multicolumn{1}{p{1.18cm}}{ \centering $[\si{km \, s^{-1}}]$} & 
\multicolumn{1}{p{1.18cm}}{ \centering $[\si{km \, s^{-1}}]$} & 
\multicolumn{1}{p{1.35cm}}{ \centering $[\si{m^{-2} \, h^{-1}}]$} \vline & 
& 
\multicolumn{1}{p{0.75cm}}{ \centering $[\si{AU}]$ } & 
\multicolumn{1}{p{0.5cm}}{ \centering $[\si{^\circ}]$ } & 
\multicolumn{1}{p{1.18cm}}{ \centering $[\si{km \, s^{-1}}]$} & 
\multicolumn{1}{p{1.18cm}}{ \centering $[\si{km \, s^{-1}}]$} & 
\multicolumn{1}{p{1.35cm}}{ \centering $[\si{m^{-2} \, h^{-1}}]$} \\
\hline
$27/10/19$ & $0.886$ & $316$ & $7.9$ & $17.7$ & $0.13$ & 
$14/06/23$ & $0.878$ & $66$ & $8.9$ & $22.5$ & $0.27$ \\
$28/11/19$ & $0.911$ & $335$ & $-5.1$ & $17.3$ & $0.12$ & 
$28/01/23$ & $0.928$ & $99$ & $-5.0$ & $21.6$ & $0.38$ \\
$19/12/18$ & $0.785$ & $306$ & $15.4$ & $19.9$ & $0.16$ & 
$16/05/22$ & $0.821$ & $41$ & $13.8$ & $25.1$ & $0.43$ \\
$20/02/19$ & $0.785$ & $348$ & $-15.5 $ & $20.2$ &$0.28$ & 
$24/08/22$ & $0.822$ & $118$ & $-13.7$ & $25.7$ & $0.89$ \\
$21/07/19$ & $0.779$ & $348$ & $-15.8 $ & $20.4$ & $0.34$ & 
$06/02/22$ & $0.818$ & $118$ & $-13.9$ & $25.9$ & $0.83$ \\
$15/05/19$ & $0.760$ & $304$ & $16.9 $ & $20.6$ & $0.23$ & 
$29/11/22$ & $0.785$ & $51$ & $13.8$ & $25.1$ & $0.48$ \\
\hline
\end{tabular}
\end{table*}

\section{Description of the model}   \label{ch:model}

In this section, we present a parametric model for bound dust detection rate on a spacecraft, which we will use to explain some of the observed features of the impact rate recorded on PSP. The model is built using the formalism of 6D distribution function over space $\vec{r}=(x,y,z)$ and velocity $\vec{v}=(v_x,v_y,v_z)$: $f(\vec{r},\vec{v}) = f(x,y,z,v_x,v_y,v_z)$, similarly to how this is done in plasma theory. A single population of bound dust is assumed. The dust number density $n(\vec{r})$ is evaluated as
\begin{equation}
    n(\vec{r}) = \iiint_{\mathbb{R}^3} f(x,y,z,v_x,v_y,v_z) \,dv_x\,dv_y\,dv_z,
\end{equation}
where the unit of $n$ is $[n]=m^{-3}$. The flux $j(\vec{r})$ as measured by the spacecraft is evaluated as the first moment of relative speed between the spacecraft and the dust. For example, the flux through a stationary test loop oriented perpendicular to the x-axis is 
\begin{equation}
    j_x(\vec{r}) = \iiint_{\mathbb{R}^3} |v_x| f(x,y,z,v_x,v_y,v_z) \,dv_x\,dv_y\,dv_z, \label{eq:f_moments}
\end{equation}
where the unit of $j_x$ is $[n]=m^{-3}s^{-1}$. The model in this form does not include the distribution of masses, and merely assumes all the dust grains are detected on contact with the spacecraft, regardless of the impact speed $v_{impact}$ or grain's mass $m$. The focus of the model is not to explain the absolute amount of the detected dust, the model works with a multiplicative prefactor. 

The spatial scaling of density $n(\vec{r})$ is assumed spherically symmetric:
\begin{equation}
    n(\vec{r}) \propto r^\gamma,
\end{equation}
where $r$ is the distance from the Sun. This allows for the presented study of the exponent $\gamma$. 

An important component of the model is the $6D$ distribution function $f$, which describes the dust cloud and is derived in Appendix \ref{app:3d_density} to be in the shape
\begin{equation}
    f(x,y,z,v_x,v_y,v_z) = C \cdot (rv_\phi)^\gamma \delta(z)\delta(v_z)\delta\left( v_r \pm \Tilde{v} \right),
\end{equation}
where $\Tilde{v}$ is the radial speed of dust given by
\begin{equation}
    \Tilde{v} = \frac{\sqrt{(e^2-1) \mu^2 + 2 \mu v_\phi^2 r - v_\phi^4 r^2}}{v_\phi r},
\end{equation}
which is Eq.~\ref{eq:derivation_6d_density} of Appendix \ref{app:3d_density}. The independent integration variable for the moments of $f$ (such as Eq.~\ref{eq:f_moments}) is chosen to be the dust azimuthal speed $v_\phi$. The integration boundaries are then the lower-most and the higher-most speeds the dust grains might have, given their eccentricity $e$ and the effective gravity $\mu(\beta)$. Therefore, 
\begin{equation}
    \sqrt{\frac{(1 - e)\mu}{r}} < v_\phi < \sqrt{\frac{(1 + e)\mu}{r}},
\end{equation}
which is Eq.~\ref{eq:derivation_integration_boundaries} of Appendix \ref{app:3d_density}. 

The model captures dust's eccentricity $e$, inclination $\theta$, radiation pressure to gravity ratio $\beta$, and the fraction of retrograde dust grains in the population $rp$. The tilt of the dust cloud is not included, as it is not higher than a few degrees \citep{mann2006dust} and therefore inconsequential for the current effort. The model can be generalized to out of ecliptics case by assuming dependence on the distance from the plane of ecliptics $z$. Yet, the spacecraft of interest operate very close to the ecliptic plane and the current assumption is deemed sufficient for the present study. The model is capable of capturing dependence of the flux through the surface $i$, denoted $j_i$ on impact speed by evaluating a moment different from $|v_{impact}|$. In this work, we assume the dependence
\begin{equation}
    j_i \propto v_{impact}^\epsilon, \label{eq:velocity_scaling}
\end{equation}
where $\epsilon$ is the relative speed exponent, equivalent to $1+\alpha \delta$ as used in several publications \citep{szalay2021collisional,zaslavsky2021first,kociscak2023modeling}, where $\alpha$ is the proportionality exponent in the charge generation equation
\begin{equation}
    q \propto m v^\alpha,
\end{equation}
and $\delta$ is the slope of the mass distribution
\begin{equation}
    n(m) \propto m^{-\delta}. \label{eq:mass_distribution}
\end{equation} 
The model treats all the parameters $e;\theta;\beta;rp;\gamma;\epsilon$ as single values (degenerate distributions). This can be, owing to the linearity, generalized to a sum of terms approximating an arbitrary distribution of these parameters, if desired. The model might in principle deal with an arbitrary shape of the spacecraft, but since the spacecraft of interest is PSP, a cylindrical shape with the axis pointing towards the Sun is assumed. 

Practically, the model evaluates the flux on the spacecraft, given the position and the speed of the spacecraft, and is therefore straight-forward to use the model to generate flux profiles starting with a position table of the spacecraft of interest, which is presently PSP. 

The derivation of the equations and a detailed discussion of the model is in Appendix \ref{app:3d_density}. Namely: in Sec.~\ref{ch:a:phase_space_density}, the assumptions on the distribution function $f$ are explained, and in Sec.~\ref{ch:a:orbital}, the orbital dynamics equations are laid out, which are needed to integrate the flux. The integration is done in Sec.~\ref{ch:a:integration} and normalization of the flux to a known number density at $\SI{1}{AU}$ is explained in Sec.~\ref{ch:a:normalization}. Sec.~\ref{ch:a:number_density_validity} examines the assumption of power-law scaling of perihelia dust density, which is an important assumption for the derivation. The model from Sec.~\ref{ch:a:phase_space_density} -- \ref{ch:a:number_density_validity} includes the free parameters $e;\beta;\gamma$, and in Sec.~\ref{ch:a:generalization} we generalize the model to account for the remaining parameters $\theta;rp;\epsilon$. 

\section{Observed flux and model results} \label{ch:results}

In this section, we compare the post-perihelion data of each orbit with the results of the model for orbital parameters representing each of the encounter groups. These are described in Appendix \ref{app:ephemeris}. We choose the post-perihelion region for two reasons. First, the bound dust impacts are more frequent than those of $\beta$-meteoroids in this region.
Second, if the sunward side of PSP is less sensitive to dust impacts, this matters the least in post-perihelia, since the sunward side is less exposed. We study post-perihelia outward of $\SI{0.15}{AU}$, since inward the dust detection process might change, as the properties of PSP change close to the Sun, as well as to avoid the possible dust depletion zone. In the next step, we also study the near-perihelion minimum of flux observed in the data to see, which features of the dust cloud included in the model might cause the dip. 

\subsection{Scaling of the flux with distance} \label{ch:scaling}

One of the features which the model should reproduce is the scaling of the observed flux with the heliocentric distance. In this section, we compare the model results to the data in the region between $\SI{0.15}{AU}$ and $\SI{0.5}{AU}$, where the flux is dominated by bound dust impacts.

\begin{figure}[h]
 	\centering
 	\includegraphics[width=9cm]{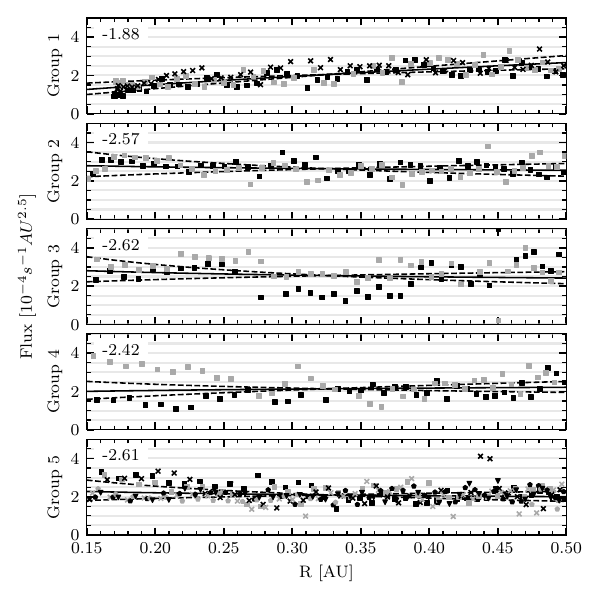}
 	\caption{The flux as detected by PSP in the outbound part of each orbit, compensated by $R^{-2.5}$, grouped by orbital groups. Individual encounters within a group are distinguished by markers of different shape or color. The solid lines are the power-law least squares fits to the observed flux, exponent of which is shown in the top left corner of each panel. To demonstrate the approximate accuracy of the $R^{-2.5}$ scaling, the dashed lines are the power-law fits of the data, with the exponent offset by $\pm 0.3$ from the least squares fit.}
 	\label{fig:compensted_flux}
\end{figure}

Inspection of the data shows that, with the exception of the first orbital group, which has its perihelion outside of $\SI{0.15}{AU}$, the flux scales approximately as $j\propto R^{-2.5}$ over the outbound leg of each orbit. This is shown in Fig.~\ref{fig:compensted_flux}. A variation is observed between individual orbits within the same orbital group, which was previously attributed to the stochastic nature of the dust cloud \citep{malaspina2020situ}. 

\begin{figure}[h]
 	\centering
 	\includegraphics[width=9cm]{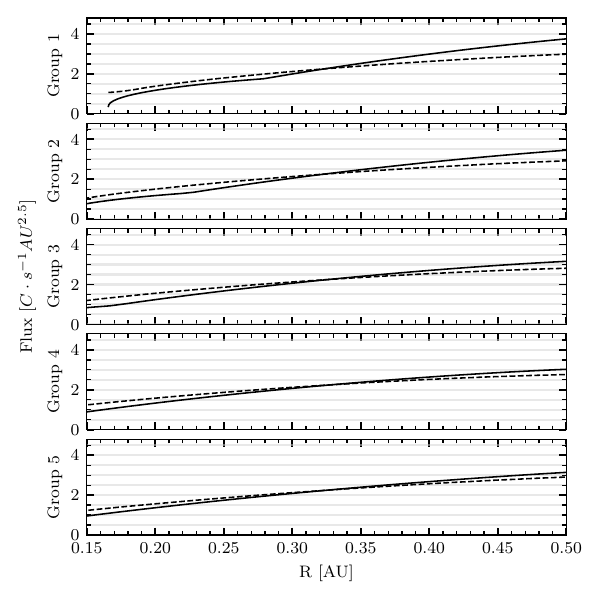}
 	\caption{The base model-predicted flux is shown in the solid line, compensated by $R^{-2.5}$. The slope is considerably shallower than $\propto R^{-2.5}$, hence the inclining trend after the compensation. The model-predicted flux in the case of $e=0.5; \theta=45^\circ{}; rp=10\%; \beta=0.5; \gamma=-1.3; \epsilon=1$ is shown in the dashed line. Even these, rather extreme, assumptions don't suffice to explain the slope observed in the data.}
 	\label{fig:compensated_insufficient_model}
\end{figure}

The base model is considered: $e=0; \theta=0; rp=0; \beta=0; \gamma=-1.3; \epsilon=1$. This is in line with the assumption of particles on circular orbits, with no inclination, no radiation pressure, without retrograde grains, and with spatial number density scaling as $n \propto R^{-1.3}$ and the assumption that every grain is always detected, if impact happened: $j \propto v_{impact}$. It is found that this assumption in not compatible (Fig.~\ref{fig:compensated_insufficient_model}) with the slope observed in the data (Fig.~\ref{fig:compensted_flux}), since the dependence produced by the model is appreciably shallower than $\propto R^{-2.5}$.

\begin{figure}[h]
 	\centering
 	\includegraphics[width=9cm]{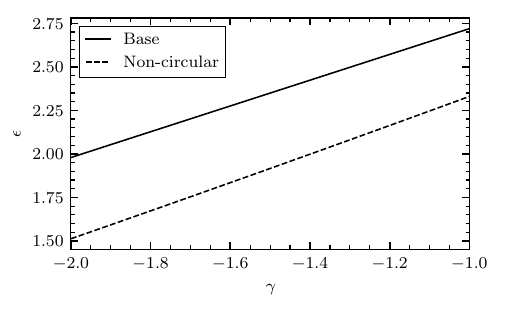}
 	\caption{The combinations of $\gamma$ and $\epsilon$ which produce a profile approximately matching the slope of $-2.5$ in the region of $\SI{0.15}{AU} < R < \SI{0.5}{AU}$. The base model assumes $e=0; \theta=0; rp=0; \beta=0$ and the non-circular model assumes rather extreme $e=0.5; \theta=45^\circ{}; rp=0.1; \beta=0.5$.}
 	\label{fig:gamma_epsilon}
\end{figure}

\begin{figure}[h]
 	\centering
 	\includegraphics[width=9cm]{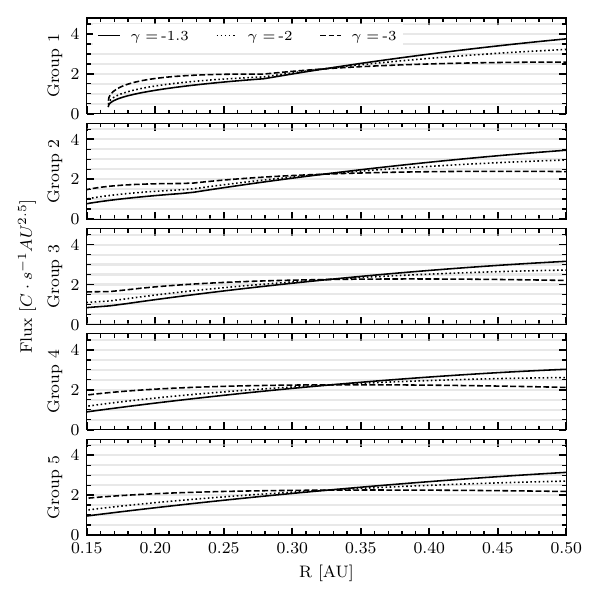}
 	\caption{The base model-predicted flux is shown in the solid line. In addition, the influence of the heliocentric distance exponent $\gamma$ on the slope is demonstrated.}
 	\label{fig:compensated_gamma}
\end{figure}

\begin{figure}[h]
 	\centering
 	\includegraphics[width=9cm]{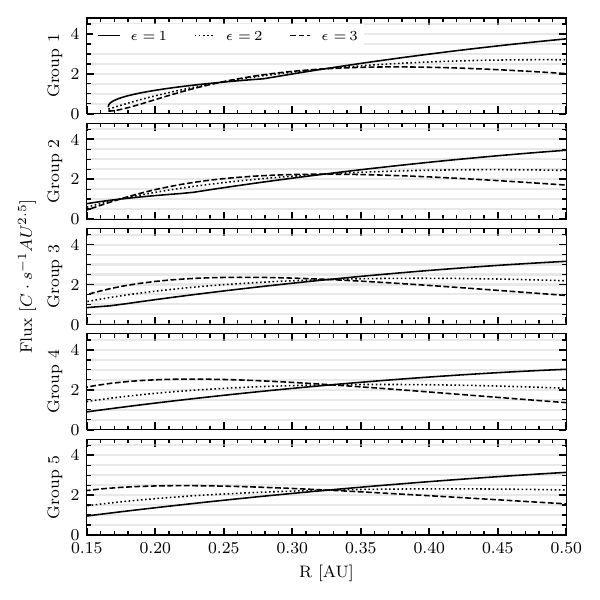}
 	\caption{The base model-predicted flux is shown in the solid line. In addition, the influence of the velocity exponent $\epsilon$ on the slope is demonstrated.}
 	\label{fig:compensated_epsilon}
\end{figure}

\begin{figure}[h]
 	\centering
 	\includegraphics[width=9cm]{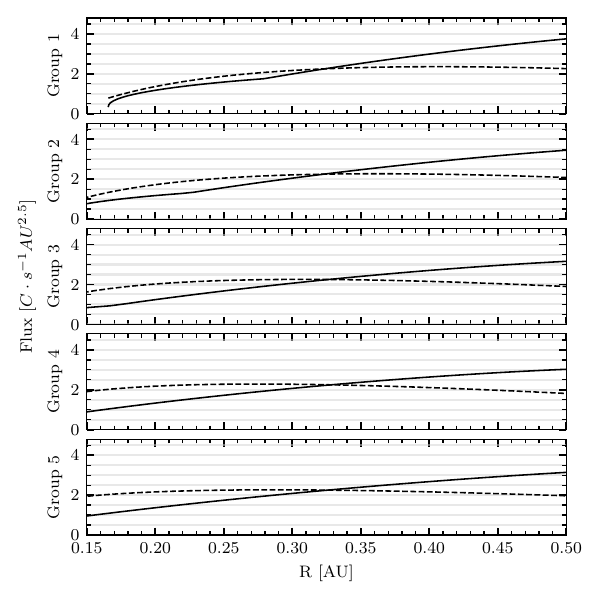}
 	\caption{The base model-predicted flux is shown in the solid line. In addition, the model with parameters:  $e=0.1; \theta=10^\circ{}; rp=0.03; \beta=0.05; \gamma=-1.9; \epsilon=2$ is shown as a representative of a viable option.}
 	\label{fig:compensated_viable}
\end{figure}

We study, what combination of parameters changes the slope to the desired $\propto R^{-2.5}$. It is found that the parameters $e; \theta; \beta; rp$ all influence the slope in the desired direction (see Appendix \ref{app:individual_parameters_slope}), yet even in the most favorable case, they do not suffice to explain the slope observed in the data, as is shown in Fig.~\ref{fig:compensated_insufficient_model}. It is also seen from Fig.~\ref{fig:compensated_insufficient_model} that especially the flux during the later orbits is very little influenced by these four parameters. The explanation, therefore, lies at least partially in the scaling of density with the two parameters not yet varied: the heliocentric distance exponent $\gamma$ and the relative speed exponent $\epsilon$. These both influence the slope appreciably, as is seen in Figs.~\ref{fig:compensated_gamma} and \ref{fig:compensated_epsilon}. We see that in the case of all the other parameters being equal to the base value, $\gamma \approx 3$, resp. $\epsilon \approx -3$ show nearly flat plots, and, therefore, serve as upper estimates of the values. Many combinations of the six parameters are capable of reproducing the right slope. We therefore study the range of combinations, which reconstructs the slope acceptably well. A viable combination of parameters is shown in Fig.~\ref{fig:compensated_viable}, but we note that the slope is not very sensitive to changes in $e; \theta; \beta; rp$. Viable combinations of the most influential parameters: $\gamma$ and $\epsilon$ are shown in Fig.~\ref{fig:gamma_epsilon}, where the other four parameters are included together in two cases: the base case, and the upper estimate: the dashed case from Fig.~\ref{fig:compensated_insufficient_model}. Fig.~\ref{fig:gamma_epsilon} shows $-2 < \gamma < -1$, which is the range of expected values of $\gamma$ for bound dust \citep{ishimoto1998modeling}. We find our results compatible with previously reported $\gamma \approx 1.3$ \citep{leinert1981zodiacal,stenborg2021psp}, in which case $2 < \epsilon < 2.5$

\subsection{Near-sun profile} \label{ch:perihelia}

The model is capable of reproducing features observed in the data close to perihelia. Notably, there is a distinct minimum in the measured flux in the perihelion. This minimum was attributed to the velocity alignment between PSP and bound dust \citep{szalay2021collisional}, which is implicitly included in the present model as well. In this section, the near-solar flux is studied as a function of the free parameters of the model. 

\begin{figure}[h]
 	\centering
 	\includegraphics[width=9cm]{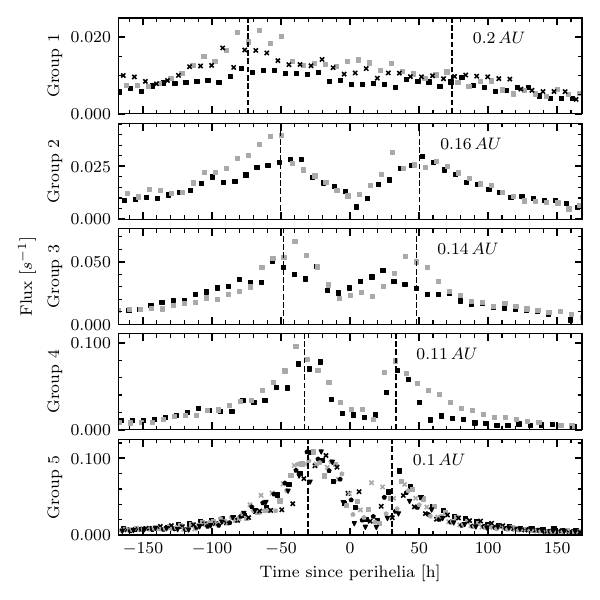}
 	\caption{The detection rate of dust impacts near the perihelia. The data from individual solar encounters are grouped according to the orbital groups. Individual encounters within a group are distinguished by markers of different shape or color.}
 	\label{fig:near_perihelia_data}
\end{figure}

The observational data are shown in Fig.~\ref{fig:near_perihelia_data} and the near-perihelion minimum is apparent. We note that the flux is not symmetric around perihelia due to the presence of $\beta$-meteoroids in pre-perihelia, where the relative speed between them and the spacecraft is high. We focus on the post-perihelia, as we will seek the consistence between these and the model predictions. We also note that the PSP perihelia lie well inside the $\beta$-meteoroid creation region \citep{szalay2021collisional}, where the $\beta$-meteoroid grains still have an important angular momentum and their trajectories are therefore similar to the trajectories of bound dust grains, making the distinction less clear.

\begin{figure}[h]
 	\centering
 	\includegraphics[width=9cm]{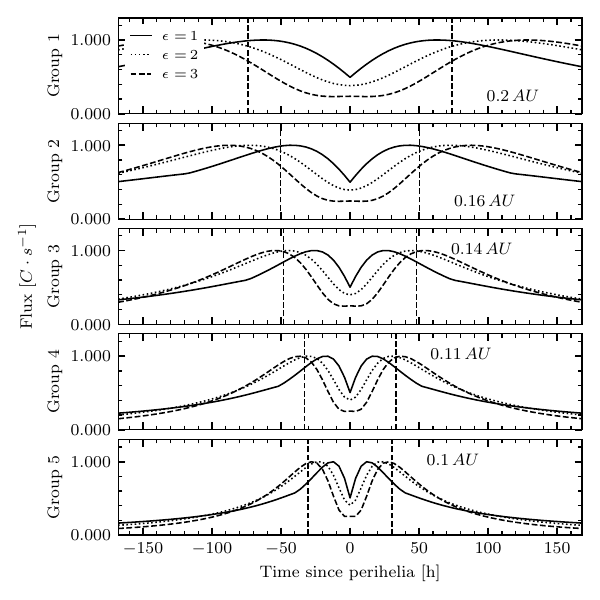}
 	\caption{The base model-predicted flux near the perihelia is shown in solid line. Different values of $\epsilon$ are shown for comparison. The same vertical dashed lines as in Fig.~\ref{fig:near_perihelia_data} are shown for reference.}
 	\label{fig:near_perihelia_model_vexp}
\end{figure}

The base model predicted flux is shown in Fig.~\ref{fig:near_perihelia_model_vexp} and we note it is symmetric around the perihelia: since only the bound dust population is assumed, and the heat shield is assumed as sensitive as the rest of the spacecraft, there is nothing to cause the asymmetry. The same set of vertical lines, approximately corresponding to the heliocentric locations of the maxima are shown symmetric around the perihelia in Figs.~\ref{fig:near_perihelia_data} and \ref{fig:near_perihelia_model_vexp}. There is no post-perihelion maximum in orbital group $1$, the vertical line is based solely on the pre-perihelion maximum. In case of the base model (as before, $e=0; \theta=0; rp=0; \beta=0; \gamma=-1.3; \epsilon=1$), the maxima in the flux are predicted decidedly closer to the perihelia than observed. It is observed in the same figure that by varying the velocity exponent $\epsilon$ the location of the expected maxima is moved, possibly to the extent that it is consistent with the data. None of the other parameters influences the location of the maxima of the flux appreciably and they are shown and discussed in App.~\ref{app:individual_parameters_perihelia}. They however do influence the relative depth of the near-perihelion flux minimum. Increasing any value except for the velocity exponent $\epsilon$ would result in a shallower perihelion dip, see the comparison on the influence of each individual parameter in App.~\ref{app:individual_parameters_perihelia}. We note that the dip is predicted shallower (approx. $\SI{50}{\%}$ of the maximum) than observed (less than $\SI{25}{\%}$ of the maximum) even in the base case.

\begin{figure}[h]
 	\centering
 	\includegraphics[width=9cm]{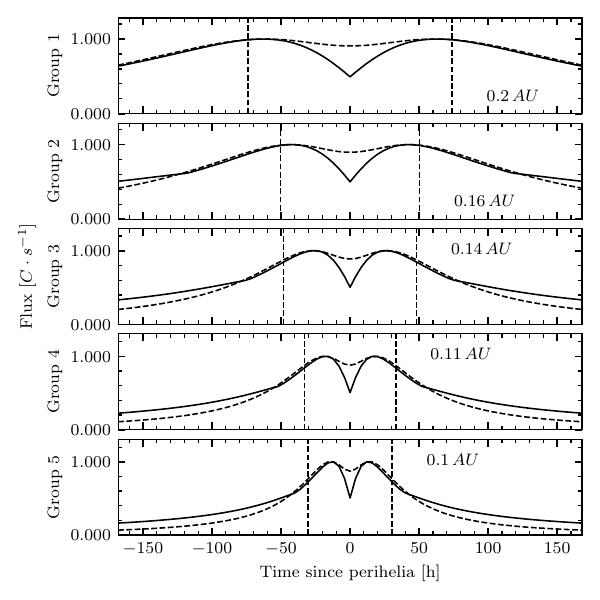}
 	\caption{The base model-predicted flux near the perihelia is shown in solid line. In addition, the model with parameters:  $e=0.1; \theta=10^\circ{}; rp=0.03; \beta=0.05; \gamma=-1.9; \epsilon=2$ is shown as a representative of a viable option.}
 	\label{fig:near_perihelia_model_viable}
\end{figure}

Fig.~\ref{fig:near_perihelia_model_viable} shows the same combination of parameters as Fig.~\ref{fig:compensated_viable} does, which was found reasonable and viable to explain the observed post-perihelion slope. Even with this reasonably conservative parameter choice, the perihelion dip is a lot less pronounced, due to a less sharp alignment between the spacecraft's and the dust's speed. It is also observed that $\epsilon$ is less effective at changing the position of the maxima, if other parameters are higher than in the base model. Therefore, we find it unlikely that the near perihelion dip is solely due to the velocity alignment between the dust cloud and the spacecraft.

\section{Discussion} \label{ch:discussion}

To extract physical information from the dust counts data of PSP and SolO, we performed different analyses in different heliospheric regions. In this section, we discuss our results with respect to the heliocentric distance. 

\subsection{Near $\SI{1}{AU}$}

Near $\SI{1}{AU}$, the comparison between PSP and SolO is possible and shows that PSP detects fewer dust impacts compared to SolO (see Fig.~\ref{fig:case_study}), especially in pre-perihelia. This is possibly an instrumental effect of a smaller sized dust being detected more effectively with SolO. As PSP's sunward side, where the heat shield is, has a different surface material to the rest of the spacecraft, it is possibly less sensitive to dust impacts, compared to the rest of the spacecraft (see Fig.~\ref{fig:charge_yield}). This implies that $\beta$-meteoroid impacts might be underrepresented in the data, as these are more likely to impact on the heat shield side, compared to the bound dust grains. Other instrumental effects might play a role, such as the position of antennas, and/or the presence of solar panels.

Beyond the likely instrumental cause of the difference of the flux on PSP and SolO, physical explanations are possible. Small sub-micron dust with $r<\SI{100}{nm}$ might be influenced by electromagnetic force. It was shown for dust of radius $r \leq \SI{30}{nm}$ that the flux might change both with solar cycle, that is on the order of years \citep{poppe2022effects}, and with solar rotation, that is on the order of days \cite{poppe2020effects,mann2021dust} even in the case of a symmetric source. A similar, albeit weaker effect might play a role for $r\approx\SI{100}{nm}$ dust as well. The individual points in space where we compared PSP and SolO (Fig.~\ref{fig:case_study}) are separated by months or weeks for a given spacecraft, and pre-perihelion and post-perihelion data points alternate in time. Therefore, a possible long-term change from 2019 to 2022 does not strongly affect the result. Due to the low number of data points, we can not dismiss the possible influence of a well-timed short-term variation, which needs to be on the order of $\SI{25}{\%}$ to explain the observed difference. With the low number of points, a stochastic counting error is not negligible, but the points are based on tens and hundreds of detections over $14$ days each, leading to error bars smaller than the observed variance.

We note that the points of similar heliocentric distance and speed between PSP and SolO lie on different heliocentric longitudes. Localized sources of $\beta$-meteoroids \citep{szalay2021collisional} or the interstellar dust \citep{mann2010interstellar} might contribute to the difference observed between the spacecraft, although there are no indications of localized $\beta$-meteoroid sources at $\SI{1}{AU}$. The perihelia of SolO in $2022$ and $2023$ are oriented close to the upstream direction of the interstellar dust, which means that more interstellar dust is likely present in the pre-perihelion data than in the post-perihelion data. The interstellar dust flux $f_{ISD}$ was reported to be coming from the heliocentric longitude of approximately $\lambda \approx 250 \, \si{^\circ}$ with the flux of $f_{ISD} < 0.036 \, \si{m^{-2} s^{-1}}$ between 2016 and 2020 \citep{babic2022phd}. This is an order of magnitude lower than the pre-perihelia SolO fluxes reported here (Tab.~\ref{tab:encounter_paramters}). Although the interstellar dust might contribute to the difference, its flux is too low to explain the whole difference. 

\subsection{Between $\SI{0.15}{AU}$ and $\SI{0.5}{AU}$}

Between $\SI{0.15}{AU}$ and $\SI{0.5}{AU}$, we were looking for the solutions of a parametric bound dust model, which would explain the observed slope of the detected flux. The two lines of $\epsilon(\gamma)$ in Fig.~\ref{fig:gamma_epsilon} may be regarded as an upper and a lower estimate of the velocity exponent $\epsilon$. We find that $\epsilon$ lies between $1.5$ and $2.7$, assuming the value of the exponent $\gamma$ between $-2$ and $-1$. This is much lower then previously assumed $4.15$ \citep{szalay2021collisional}. We note that the bound dust which contributes to the results is mostly of the $\si{\mu m}$ and sub-$\si{\mu m}$ size. For a reasonably conservative $\alpha \approx 3.5$ \citep{collette2014micrometeoroid} in $\epsilon = 1 + \alpha \delta$, it is implied that $0.14 < \delta < 0.49$, which is a much lower value than $\delta \approx 0.9$ observed for larger dust \citep{grun1985collisional,pokorny2024long}. The value of $\alpha$ is better constrained than the value of $\delta$. This is therefore an indication that the mass distribution may not be a single power-law in the mass range of interest and at all heliocentric distances. However, this makes sense, because of the narrow mass interval: the power-law distribution of masses was described over $20$ decades of magnitude in mass \citep{grun1985collisional}, while PSP detects grains wihch span less than two orders of magnitude \citep{malaspina2023dust}. Since the dust in question is bound, there is a necessary depletion in the small size region, as small dust ($r \lesssim \SI{100}{nm}$) is not bound, due to high $\beta$. Interestingly, our value is compatible with $\delta \approx 0.34$, which was reported by \cite{zaslavsky2021first} for $\beta$-meteoroids, which are also limited in mass, and the power-law distribution of masses is therefore also problematic.

\subsection{Near perihelia}

We studied the compatibility between the bound dust model and the observed perihelion dip. A similar dip is formed due to the velocity alignment between the spacecraft and the bound dust, but the dip is too shallow and readily smeared out by non-zero eccentricity, inclination, or other parameters. If the near perihelion dip is not due to the velocity alignment, other factors might contribute to the dip, and we list several of them. First, the dust number density is believed to diminish closer to the Sun. A dust free zone was hypothesised to exist \citep{russell1929meteoric}, since the dust grains do not survive for long in the extreme conditions near the Sun. It was estimated using WISPR that a dust depletion zone enveloping the dust free zone lies inward of $19 \, R_{sun} \approx \SI{0.09}{AU}$ and the dust free zone likely lies inward of $5 \, R_{sun} \approx \SI{0.023}{AU}$ \citep{stenborg2021psp}. Such dust depletion zone may explain the apparent depletion of dust near perihelia. The second possible explanation of the dip is the ineffective detection. The antenna detection process is very much dependent on the spacecraft's charge state and the surrounding environment \citep{shen2021electrostatic,babic2022analytical,shen2023variability}. An indication that the process changes rapidly inward of $\SI{0.15}{AU}$ is that the spacecraft's potential seems to be a steep function of the heliocentric distance in this region, potentially disturbing the otherwise effective dust detection. For example, the heat shield becoming conductively coupled to the spacecraft's body inward of $\approx \SI{0.16}{AU}$ \citep{diaz2021parker} also changes spacecraft's charge state. See Sec.~\ref{sec:potential} for the discussion of the spacecraft potential. The third possible contribution to the dip is that if the relative speed between the dust and the spacecraft is higher than a certain threshold, all the bound dust grains are detected, and therefore the flux plateaus. This would not produce the dip on its own, but would lead to a shallower growth near the Sun, compared to the case, when the proportionality of flux to $\propto v^\epsilon$ is assumed all the way. In addition to these three explanations, we note that the minima near the perihelia might be a result of maxima before and after the perihelia, rather then a depletion. Such maxima might result from either crossing the Geminids $\beta$-stream \citep{szalay2021collisional}, or the spacecraft getting more sensitive to dust impacts, possibly due to a higher potential (Fig.~\ref{fig:potentials}) and/or better sensitivity of the heat shield. 

\subsection{Further work}

We developed a model to describe bound dust impact rates onto spacecraft, which work with sharp values of the free parameters, but is easily generalized to distributions. Other free parameters are feasible to be included, such as the tilt of the dust cloud with respect to the ecliptic plane, which might be useful for modelling the flux once the orbit of SolO becomes more inclined. It is straight-forward to develop a similar model for dust on hyperbolic trajectories, such as $\beta$-meteoroids or interstellar dust. The distribution of inclinations and eccentricities within $\beta$-meteoroid cloud is worthy of future investigation. 

The masses are presently treated in a crude way, assuming power-law distribution of masses, which translates to the efficiency of detection. As we argue, the assumption of a power-law distributed masses might not be justified. A $7D$ distribution function describing masses, in addition to the phase space, would offer a more nuanced model, but would push the limits of what information is possibly retrieved from the data. 

\section{Conclusions} \label{ch:conclusion}

The impact rate model, which we developed in this work, takes into account dust cloud parameters: eccentricity $e$, inclination $i$, radiation pressure to gravity ratio $\beta$, and retrograde dust fraction $rp$. In addition, the model takes into account semi-empirical parameters: the exponent $\gamma$ of number density dependence on heliocentric distance and the exponent $\epsilon$ of the detection rate dependence on the impact speed. We compared the model results to the dust count data of the first $16$ orbits of PSP.

Although the model does produce a dip in flux due to the velocity alignment between the dust and the spacecraft in the circular dust base case, but the dip is not sufficient and is smeared away easily, especially with non-zero eccentricity. The dip observed close to each of the perihelia for the third and subsequent orbits of PSP is reproduced neither as deep nor as wide by the model as is observed in the data. Therefore, other effects contribute to the dip beyond the velocity alignment, namely the dust depletion zone or instrumental effects. 

The parameters of the dust cloud: $e,i,\beta,rp$ all have minor influence on the model profile, while the semi-empirical parameters $\gamma$ and $\epsilon$ are crucial. By varying these, we can reproduce the observed dependence of flux in the post-perihelion region on the heliocentric distance between $\SI{0.15}{AU}$ and $\SI{0.5}{AU}$, where the influence of $\beta$ dust is expected the lowest. The parameter $\epsilon$, which represents the combined influence of dust mass distribution and impact charge production, is found lower than what was previously used for PSP dust data analysis, and likely between $1.5$ and $2.7$. This is consistent with the slope $\delta$ of differential mass distribution of $\si{\mu m}$ and sub-$\si{\mu m}$ dust between $0.14$ and $0.49$, which is shallower than what was reported for bigger bound dust further away from the Sun. 

A comparison of the dust counts of PSP and SolO shows that PSP observed comparatively less dust in pre-perihelia than in post-perihelia, with the difference of about $\SI{25}{\%}$. This suggests that PSP's sun-facing side, and therefore the heat shield (TPS), offers a less dust sensitive target, compared to the other surfaces of PSP. Because of this instrumental effect, the observations are not incompatible with a stationary and symmetric dust cloud. However, due to the low number of data points, we can not reject the possibility that the effect is physical, possibly attributed to the short-term temporal or spatial variation of the cloud. 

\begin{acknowledgements}


Author contributions: Concept: SK, AT, IM. Model development: SK, AT. Data analysis: SK, AT. Interpretation: SK, IM, AT. Manuscript preparation: SK, IM. 

SK and AT were supported by the Tromsø Research Foundation under the grant 19\_SG\_AT.

This work on dust observations in the inner heliosphere received supported from the Research Council of Norway (grant number 262941 and 275503).

SK appreciates the constructive discussions with Jakub Vaverka, Libor Nouzák, Jamey Szalay, Mitchell Shen, and Arnaud Zaslavsky, and the assistance of David Malaspina with interpreting the PSP dust data.

The code and the data set used in this work are publicly available at\\ \url{https://zenodo.org/records/13284890}.

      
\end{acknowledgements}

\newpage

\bibliographystyle{aa}
\renewcommand{\bibname}{References}
\bibliography{refs}

\begin{appendix} 

\section{Integrating the phase-space density} \label{app:3d_density}

\subsection{Phase-space density} \label{ch:a:phase_space_density}

Assume there is a time-invariant dust density $f$ in the usual 6D phase space:
\begin{equation}
    f(\vec{r},\vec{v}) =  f(x,y,z,v_x,v_y,v_z),
\end{equation}
which is normalized to number density as 
\begin{equation}
    n(x,y,z) = \iiint_{\mathbb{R}^3} f(x,y,z,v_x,v_y,v_z) \,dv_x\,dv_y\,dv_z,
    \label{eq:number_density}
\end{equation}
which is a very useful way of looking at it, since $n(\vec{r})$ can be measured remotely for bigger ($\gtrsim \SI{1}{\mu m}$) dust grains. We note that we disregarded grain size for now, the density $n$ represents the number density of \textit{suitable} dust grains, whatever the suitability criteria are. 

Since the Solar system near the ecliptic is our main goal, we will introduce simplifying assumptions on $f$:
\begin{itemize}
    \item We assume that the plane $x \otimes y$ is the ecliptic, with $(0,0)$ point being the Sun and assume that all the grains within the distribution move within this plane, with no pole-ward component of the speed $v_z$. 
    \item We assume that the dust cloud has a rotational symmetry around the $z$ axis, and for convenience we are going to use the density $\tilde{f}$ expressed using $(r,\phi)$ instead of $(x,y)$, where $r=\sqrt{x^2 + y^2}$ and $\phi$ is the angle of rotation around $z$-axis, measured from an arbitrary ray in the ecliptic.
    \item We assume the dust grains don't collide. Then we make use of  Liouville's theorem on the space $(\vec{r},\vec{v})$.
\end{itemize}
The first assumption is translated to $f$ using degenerate distributions $\delta(\cdot)$ as
\begin{equation}
    f(x,y,z,v_x,v_y,v_z) = f(x,y,0,v_x,v_y,0) \delta_0(z) \delta_0(v_z).
    \label{eq:ecliptic}
\end{equation}
The second assumption is translated using $r,\phi$ as
\begin{equation}\begin{split}
      &f(x,y,0,v_x,v_y,0) \delta_0(z) \delta_0(v_z) \\
    = &f(r,0,0,v_x,v_y,0) \delta_0(z) \delta_0(v_z) \\
    = &\tilde{f}(r,0,0,v_x,v_y,0) \delta_0(z) \delta_0(v_z),
\end{split}\end{equation}
where the first three arguments of $f$ are all position arguments, where $\tilde{f}$ has a position, angle, and a position arguments. We also use a more compact 3D distribution $f$, with the meaning
\begin{equation}\begin{split}
    &f(x,y,z,v_x,v_y,v_z) \\
    &= \tilde{f}(r,\phi,0,v_r,v_\phi,0) \delta_0(z) \delta_0(v_z) \\ 
    &\equiv f(r,v_r,v_\phi) \delta_0(z) \delta_0(v_z) \ \forall \phi \in \mathbb{R}, \label{eq:compact_f}
\end{split}\end{equation}
since we assumed rotational symmetry in $\phi$.
The third assumption has the form of
\begin{equation}\begin{split}
    f(\vec{r}_1,\vec{v}_1) &= f(\vec{r}_2,\vec{v}_2) \Leftrightarrow \\
    f(r_1,v_{r,1},v_{\phi,1}) &= f(r_2,v_{r,2},v_{\phi,2})
    \label{eq:liouville}
\end{split}\end{equation}
provided that the points $(\vec{r}_1,\vec{v}_1),(\vec{r}_2,\vec{v}_2)$ (or, alternatively expressed points $(r_1,v_{r,1},v_{\phi,1}),(r_2,v_{r,2},v_{\phi,2})$) share the same trajectory of the system in the phase space. We note that we use velocity, not the momentum, which is justified, since we assume the mass conservation $dm/dt=0$ for each particle. We assume the dust cloud is composed of bound dust grains, each of them on a heliocentric orbit. Then Eq.~\ref{eq:liouville} holds for any two points of an orbit of a grain. If we assume all the grains follow the same gravity field with the effective gravitational parameter 
\begin{equation}
    \mu = (1-\beta) \kappa M_{Sun},
\end{equation}
where $\beta$ is the grain's radiation pressure to gravity ratio and $\kappa M_{Sun}$ is the solar gravitational parameter. Then all the grains which acquire the state of $(\vec{r}_1,\vec{v}_1)$ will also acquire the state of $(\vec{r}_2,\vec{v}_2)$, if this a valid solution for one of them. Therefore, we may study $f$ in a convenient point of the orbit of our choice while being assured, it remains the same throughout the orbit. We are soon going to see that the perihelion of the orbit of a dust grains is a convenient point. 

\subsection{Orbital mechanics} \label{ch:a:orbital}

To study the density $f$ in a point of the orbit of our choice, we must describe the orbits and be able to translate between the points within the orbit. From Eq.~\ref{eq:liouville} we know that 
\begin{equation}
    f(r_{peri},0,v_{peri}) = f(r,v_r,v_\phi),
    \label{eq:contraction_to_peri}
\end{equation}
provided that the spacecraft state in perihelion $(r_{peri},0,v_{peri})$ shares the same orbit with a general state $(r,v_r,v_\phi)$. By applying the laws of orbital motion, we are going to find the relationship between the points $(r_{peri},0,v_{peri})$ and $(r,v_r,v_\phi)$. We know the angular momentum is conserved:
\begin{equation}
    r_{peri} v_{peri} = v_\phi r \Leftrightarrow v_{peri} = \frac{v_\phi r}{r_{peri}} \Leftrightarrow r_{peri} = \frac{v_\phi r}{v_{peri}}.
    \label{eq:momentum}
\end{equation}
as well as the energy is:
\begin{equation}
    v_{peri}^2 - \frac{2\mu}{r_{peri}} = v_\phi^2 + v_r^2 - \frac{2\mu}{r}.
\end{equation}
Substituting $v_{peri}$ from the angular momentum, we get
\begin{equation}
        \left( \frac{v_\phi r}{r_{peri}} \right)^2 - \frac{2\mu}{r_{peri}} = v_\phi^2 + v_r^2 - \frac{2\mu}{r}, 
\end{equation}
and multiplying by $r^2$ we get
\begin{equation}\begin{split}
        r_{peri}^2 \left( v_\phi^2 + v_r^2 - \frac{2\mu}{r} \right) + r_{peri}(2\mu) - (v_\phi^2 r^2) &= 0 \\
        ar_{peri}^2 + br_{peri} + c &= 0.
\end{split}\end{equation}
The two formal solutions of this equation are
\begin{equation}
    r_{peri} = \frac{-b \pm \sqrt{b^2-4ac}}{2a},
\end{equation}
where $(+)$ and $(-)$ correspond to the aphelion and perihelion respectively, since we didn't assume anything other than a stationary point yet. Hence, $(-)$ corresponds to the true $r_{peri}$ and substituting back for $a,b,c$ and substituting for $v_{peri}$ from Eq.~\ref{eq:momentum} we get:
\begin{equation}\begin{split}
    r_{peri} &= \frac{-b - \sqrt{b^2-4ac}}{2a} = \frac{-\mu - \sqrt{\mu^2+\left( v_\phi^2+v_r^2-\frac{2\mu}{r} \right)(v_\phi^2r^2)}}{\left( v_\phi^2+v_r^2-\frac{2\mu}{r} \right)} \\
    v_{peri} &= \frac{v_\phi r}{r_{peri}}.
    \label{eq:peri_contraction}
\end{split}\end{equation}
These two equations are what was needed to solve Eq.~\ref{eq:contraction_to_peri}. However, our goal is to study different eccentricities. Since an arbitrary distribution of eccentricities is straight-forward to approximate with a linear combination of sharp-eccentricity terms, we will now focus on a sharp eccentricity $e$. Assuming that all the grains are not only exposed to the same effective gravity $\mu$, but they also have the same orbital eccentricity $e$, it is apparent that \textit{only one speed $v_{peri}$ is allowed in the perihelion $r_{peri}$, which conforms to the eccentricity $e$} $(\star)$. Vis-viva equation in perihelion gives
\begin{equation}
    v_{peri} = \sqrt{\mu \frac{1+e}{r_{peri}}}.
\end{equation}
Substituting for $v_{peri}$ from Eq.~\ref{eq:momentum}, we get
\begin{equation}\begin{split}
    \left(\frac{v_\phi r}{r_{peri}} \right)^2 &= \mu \frac{1+e}{r_{peri}} \\
    r_{peri} &= \frac{v_\phi^2 r^2}{\mu (1+e)}.
    \label{eq:eccentricity_bond}
\end{split}\end{equation}
This equation is the bond between an arbitrary $r,v_r,v_\phi$ and the only corresponding $r_{peri},v_{peri}$ given the eccentricity $e$. 

As a reasonable simplification \citep{giese1986three}, we assume the number density $n(x,y,z)$ in the ecliptic depends on the heliocentric distance:
\begin{equation}
    n(x,y,z) = n(r,0)\delta_0(z) = A \left( \frac{r}{r_0} \right)^\gamma \delta_0(z) = \delta_0(z) \frac{A}{r_0^\gamma} r^\gamma,
    \label{eq:radial_density}
\end{equation}
where $\gamma$ is a parameter, reasonably constrained by experiment. We normalized the expression by the number density at $r_0$, which might be for example $1 \, \si{AU}$. Assume this dependence ($\propto r^\gamma$) holds for the distribution of dust grains in their perihelia (see section \ref{ch:a:number_density_validity} for the discussion), which is surely an acceptable assumption, at least for low $e$, since at low $e$, the difference between $r_{peri}$ and $r_{aph}$ is very small. 

\subsection{Velocity moments' integration} \label{ch:a:integration} 

The net flux of particles through the $x$-plane is the first speed moment 
\begin{equation}
    j_x = \iiint_{\mathbb{R}^3} v_x f(\vec{r},\vec{v})  \,dv_x\,dv_y\,dv_z.
\end{equation}
The SI unit is $[j_x]=m^{-2}s^{-1}$. We are however not interested in the net flux $j_x$ but the total flux $j_{tot,x}$ onto the plane $x$. For the stationary plane $x$, we have
\begin{equation}\begin{split}
    j_{tot,x} &= \int_0^\infty \iint_{\mathbb{R}^2} v_x f(\vec{r},\vec{v}) \,dv_x\,dv_y\,dv_z 
    \\ &+ \int_{-\infty}^0 \iint_{\mathbb{R}^2} -v_x f(\vec{r},\vec{v}) \,dv_x\,dv_y\,dv_z.
   \end{split}\end{equation}
Should the probe be moving in $+x$-direction with the speed of $v_{p,x}$, the detected net flux is going to be
\begin{equation}\begin{split}
    j_{tot,x} &= \int_{v_{p,x}}^\infty \iint_{\mathbb{R}^2} (v_x-v_{p,x}) f(\vec{r},\vec{v}) \,dv_x\,dv_y\,dv_z 
    \\ &- \int_{-\infty}^{v_{p,x}} \iint_{\mathbb{R}^2} (v_x-v_{p,x}) f(\vec{r},\vec{v}) \,dv_x\,dv_y\,dv_z.
\end{split}\end{equation}
Since we assumed all the dust being concentrated around the ecliptic plane (Eqs.~\ref{eq:ecliptic}, \ref{eq:compact_f}):
\begin{equation}\begin{split}
    j_{tot,x} &= \int_{v_{p,x}}^\infty \iint_{\mathbb{R}^2} (v_x-v_{p,x}) f(r,v_r,v_\phi) \delta_0(z) \delta_0(v_z) \,dv_x\,dv_y\,dv_z 
    \\ &- \int_{-\infty}^{v_{p,x}} \iint_{\mathbb{R}^2} (v_x-v_{p,x}) f(r,v_r,v_\phi) \delta_0(z) \delta_0(v_z) \,dv_x\,dv_y\,dv_z
    \\
    &= \delta_0(z) \int_{v_{p,x}}^\infty \int_{\mathbb{R}} (v_x-v_{p,x}) f(r,v_r,v_\phi)  \,dv_x\,dv_y 
    \\ &- \delta_0(z) \int_{-\infty}^{v_{p,x}} \int_{\mathbb{R}} (v_x-v_{p,x}) f(r,v_r,v_\phi) \,dv_x\,dv_y.
\end{split}\end{equation}
And since we align the $x$-axis with the probe, the speed $v_x = v_r$ is the radial dust speed and $v_y = v_\phi$ is the azimuthal dust speed, both in the unit of translational speed (as not to confuse with the angular speed $\dot{\phi} \neq v_\phi = r \dot{\phi}$). The flux measured on radially oriented surfaces of the probe is
\begin{equation}\begin{split}
    j_{tot,rad} &= \delta_0(z) \int_{v_{p,rad}}^\infty \int_{\mathbb{R}} (v_r-v_{p,rad}) f(r,v_r,v_\phi)  \,dv_r\,dv_\phi
    \\ &- \delta_0(z) \int_{-\infty}^{v_{p,rad}} \int_{\mathbb{R}} (v_r-v_{p,rad}) f(r,v_r,v_\phi)  \,dv_r\,dv_\phi,
    \label{eq:general_radial}
\end{split}\end{equation}
and, analogically, the flux measured on the azimuthally oriented surfaces as
\begin{equation}\begin{split}
    j_{tot,azim} &= \delta_0(z) \int_{\mathbb{R}} \int_{v_{p,azim}}^\infty (v_\phi-v_{p,azim}) f(r,v_r,v_\phi)  \,dv_r\,dv_\phi
    \\ &- \delta_0(z) \int_{\mathbb{R}} \int_{-\infty}^{v_{p,azim}} (v_\phi-v_{p,azim}) f(r,v_r,v_\phi)  \,dv_r\,dv_\phi,
    \label{eq:general_azimuthal}
\end{split}\end{equation}
where $v_{p.azim}$ is the azimuthal speed of the probe (prograde, locally in $+y$-direction). 

To obtain the most convenient form of $f$ in an easily integrable shape, we use two additional pieces of information: 1. the dependence on the heliocentric distance (Eq.~\ref{eq:radial_density}), and 2. the fact, that Eqs.~\ref{eq:eccentricity_bond} and \ref{eq:peri_contraction} must be demanded consistent, which is the equivalent to the $(\star)$ claim. We are shortly going to integrate $f$ over $v_r$ and $v_\phi$, as in Eqs.~\ref{eq:general_radial} and \ref{eq:general_azimuthal}. We know that not all combinations of $v_r$, $v_\phi$ are possible given $r$, $e$. Instead of integrating in two dimensions, we will integrate in $v_r \otimes v_\phi$ space along the path $v_r(v_\phi)$ given by ($\star$). Hence, we relate Eqs.~\ref{eq:eccentricity_bond} and \ref{eq:peri_contraction} with the goal of obtaining $v_r(v_\phi)$:
\begin{equation}
    \frac{-\mu - \sqrt{\mu^2+\left( v_\phi^2+v_r^2-\frac{2\mu}{r} \right)(v_\phi^2r^2)}}{\left( v_\phi^2+v_r^2-\frac{2\mu}{r} \right)} = \frac{v_\phi^2 r^2}{\mu (1+e)} 
    \label{eq:bond_raw}
\end{equation}
This quadratic equation has two solutions for $v_r$:
\begin{equation}
    v_r = \pm \frac{\sqrt{(e^2-1) \mu^2 + 2 \mu v_\phi^2 r - v_\phi^4 r^2}}{v_\phi r} = \pm \Tilde{v},
    \label{eq:bond_nice}
\end{equation}
These solutions correspond to the radial speeds of pre-perihelion (in-going, $v_r<0$) and post-perihelion (out-going, $v_r>0$) dust, as at a given $r$ and with a given $v_\phi$. Since $f$ of the dust cloud is assumed stationary, the grains don´t collide and are in repetitive orbits, therefore there are exactly as many in-going as out-going. 

Now to get a convenient shape of $f$, assuming separable power-law scaling with distance (consistently with Eq.~\ref{eq:radial_density}):
\begin{equation}\begin{split}
    f(r,v_r,v_\phi) &= f(r,v_\phi)\delta\left( v_r \pm \Tilde{v} \right)
    \\ &\propto r^{\gamma}\hat{f}(v_\phi)\delta\left( v_r \pm \Tilde{v} \right),
    \label{eq:shaping_f}
\end{split}\end{equation}
where in the first step we expressed the condition \ref{eq:bond_nice} and newly used a 2D distribution $f$ according to $f(r,v_\phi)\delta\left( v_r \pm \Tilde{v} \right) = f(r,v_r,v_\phi)$ and in the second step we expressed the condition that $n(r)\propto r^{\gamma}$ using a new 1D $\hat{f}(v_\phi)$ and we lost the normalization. Since Liouville's theorem says that the density is the same along the trajectory of the grain, it is the same in perihelion as well, and its moments are the same, and therefore integrating over $z,v_z,v_r$ in an arbitrary time (LHS) and in perihelion (RHS):
\begin{equation}\begin{split}
    r^{\gamma}\hat{f}(v_\phi) &= r_{peri}^{\gamma}\hat{f}(v_{peri}) 
    \\ r^{\gamma}\hat{f}(v_\phi) &= \left( \frac{v_\phi^2 r^2}{\mu (1+e)} \right)^\gamma \hat{f}\left(\frac{\mu (1+e)}{v_\phi r}\right) \\
    r^{\gamma} \hat{f}(v_\phi) &= \left( \mu (1+e) \right)^{-\gamma} v_\phi^{2\gamma} r^{2\gamma} \hat{f}\left(\frac{\mu (1+e)}{v_\phi r}\right).
\end{split}\end{equation}
Since $\mu(1+e)$ is a plain number, and except for $\hat{f}(\dots)$ we only have $r^{c_1}$ and $v_\phi^{c_2}$ and we demand the equality for arbitrary $r,v_\phi$, the only thinkable solution for $\hat{f}$ is of the form
\begin{equation}
    \hat{f}(x) = C x^b,
\end{equation}
therefore:
\begin{equation}\begin{split}
    r^{\gamma}C v_\phi^b &= \left( \mu (1+e) \right)^{-\gamma} v_\phi^{2\gamma} r^{2\gamma} C\left(\frac{\mu (1+e)}{v_\phi r}\right)^b \\
    1 &= \left( \mu (1+e) \right)^{b-\gamma} v_\phi^{2\gamma-2b} r^{\gamma-b} \\
    1 &= \left( \frac{v_\phi^2 r}{\mu (1+e)} \right)^{\gamma-b},
\end{split}\end{equation}
Where the only suitable solution is $b=\gamma$. Therefore, $\hat{f}(x) = C x^\gamma$ Eq.~\ref{eq:shaping_f} with Eq.~\ref{eq:bond_nice} give:
\begin{equation}\begin{split}
    &f(r,v_r,v_\phi) =
    \\ &\qquad = C \cdot (rv_\phi)^\gamma \delta(z)\delta(v_z)\delta\left( v_r \pm \Tilde{v} \right)
    \\ &\qquad = C \cdot (rv_\phi)^\gamma \delta(z)\delta(v_z)\delta\left( v_r \pm \frac{\sqrt{(e^2-1) \mu^2 + 2 \mu v_\phi^2 r - v_\phi^4 r^2}}{v_\phi r} \right).
    \label{eq:derivation_6d_density}
\end{split}\end{equation}
The last parenthesis of this equation may be interpreted as the integration trajectory in the $v_r \otimes v_\phi$ space. 
Since our integration parameter of the contraction from the the $v_r \otimes v_\phi$ space to a 1D space of the path is going to be $v_\phi$, we need the integration boundaries for $v_\phi$. The integration boundaries are given by the lowermost and the uppermost $v_\phi$ the probe may encounter, given $r$,$e$. The lowest possible $v_\phi$ corresponds to the probe being in aphelion, whereas the highest corresponds to it being in the perihelion. We use Eq.~\ref{eq:bond_nice}, and both in the perihelion and in the aphelion, we get $v_r=0$, hence
\begin{equation}
    (e^2-1) \mu^2 + 2 \mu v_\phi^2 r - v_\phi^4 r^2 = 0,
\end{equation}
and this quadratic equation in $v_\phi^2$ has two solutions:
\begin{equation}
    v_\phi^2 = \frac{(1 \pm e)\mu}{r},
\end{equation}
corresponding to the highest possible $v_\phi$ (in the case $r$ is the perihelion) and the lowest possible (in the case $r$ is the aphelion). Negative $v_\phi$ would correspond to the dust grains on retrograde orbits and are disregarded as we only want to include prograde dust grains for now. These are, therefore, our integration boundaries: 
\begin{equation}
    \sqrt{\frac{(1 - e)\mu}{r}} < v_\phi < \sqrt{\frac{(1 + e)\mu}{r}}.
    \label{eq:derivation_integration_boundaries}
\end{equation}
We note that these correspond to the degenerate solutions of Eq.~\ref{eq:bond_nice}, which makes sense, since Eq.~\ref{eq:bond_nice} defines a cyclic trajectory in the $v_r \otimes v_\phi$ space. 

\paragraph{Radial flux}
\begin{equation}\begin{split}
    j_{tot,rad} &= \delta(z) \int_{v_{p,rad}}^\infty \int_{\mathbb{R}} (v_r-v_{p,rad}) f(r,v_r,v_\phi)  \,dv_r\,dv_\phi
    \\ &- \delta(z) \int_{-\infty}^{v_{p,rad}} \int_{\mathbb{R}} (v_r-v_{p,rad}) f(r,v_r,v_\phi)  \,dv_r\,dv_\phi
    \\ &= \delta(z) C r^\gamma \int_{v_{p,rad}}^\infty \int_{\mathbb{R}} (v_r-v_{p,rad}) v_\phi^\gamma \delta\left( v_r \pm \Tilde{v} \right) \,dv_r\,dv_\phi
    \\ &-  \delta(z) C r^\gamma \int_{-\infty}^{v_{p,rad}} \int_{\mathbb{R}} (v_r-v_{p,rad}) v_\phi^\gamma \delta\left( v_r \pm \Tilde{v} \right) \,dv_r\,dv_\phi.
\end{split}\end{equation}
The expression contains two terms: $j_{tot,rad} = \delta_0(z) C r^\gamma ( j_{rad}^+ - j_{rad}^- )$, which have the boundaries $v_r > v_{p,rad}$ and $v_{p,rad} > v_r$ respectively, which corresponds to flux on the sun-facing ($+$) and on the anti-sunward ($-$) respectively. We translate these boundaries from $v_r$ to $v_\phi$ in order to integrate over the parameter $v_\phi$ using the Heaviside function. Each of these two has two variants: post-perihelion (\textit{post}, $v_r>0$) and pre-perihelion (\textit{pre}, $v_r<0$) dust, as $\pm \Tilde{v}$ in Eq.~\ref{eq:bond_nice}. Thus, we get four integral terms, each with a prefactor of $1/2$:
\begin{equation}\begin{split}
    j_{rad}^{+,pre} &= \frac{1}{2} \int_{v_{p,rad}}^\infty \int_{\mathbb{R}} (v_r-v_{p,rad}) v_\phi^\gamma \delta\left( v_r + \Tilde{v} \right) \,dv_r\,dv_\phi
    \\ &= \frac{1}{2} \int_{\mathbb{R}} \int_{\mathbb{R}} (v_r-v_{p,rad}) v_\phi^\gamma \delta\left( v_r + \Tilde{v} \right) H(v_r - v_{p,rad}) \,dv_r\,dv_\phi
    \\ &= \frac{1}{2} \int_{\mathbb{R}} (-\Tilde{v}-v_{p,rad}) v_\phi^\gamma H(-\Tilde{v} - v_{p,rad}) \,dv_\phi
    \\ &= \frac{1}{2} \int_{\sqrt{\frac{(1 - e)\mu}{r}}}^{\sqrt{\frac{(1 + e)\mu}{r}}} (-\Tilde{v}-v_{p,rad}) v_\phi^\gamma H(-\Tilde{v} - v_{p,rad}) \,dv_\phi,
    \label{eq:j_rad_plus_pre}
\end{split}\end{equation}
\begin{equation}\begin{split}
    j_{rad}^{+,post} &= \frac{1}{2} \int_{v_{p,rad}}^\infty \int_{\mathbb{R}} (v_r-v_{p,rad}) v_\phi^\gamma \delta\left( v_r - \Tilde{v} \right) \,dv_r\,dv_\phi
    \\ &= \frac{1}{2} \int_{\sqrt{\frac{(1 - e)\mu}{r}}}^{\sqrt{\frac{(1 + e)\mu}{r}}} (\Tilde{v}-v_{p,rad}) v_\phi^\gamma H(\Tilde{v} - v_{p,rad}) \,dv_\phi,
    \label{eq:j_rad_plus_post}
\end{split}\end{equation}
\begin{equation}\begin{split}
    j_{rad}^{-,pre} &= \frac{1}{2} \int_{-\infty}^{v_{p,rad}} \int_{\mathbb{R}} (v_r-v_{p,rad}) v_\phi^\gamma \delta\left( v_r + \Tilde{v} \right) \,dv_r\,dv_\phi
    \\ &= \frac{1}{2} \int_{\sqrt{\frac{(1 - e)\mu}{r}}}^{\sqrt{\frac{(1 + e)\mu}{r}}} (-\Tilde{v}-v_{p,rad}) v_\phi^\gamma H(\Tilde{v} + v_{p,rad}) \,dv_\phi,
    \label{eq:j_rad_minus_pre}
\end{split}\end{equation}
\begin{equation}\begin{split}
    j_{rad}^{-,post} &= \frac{1}{2} \int_{-\infty}^{v_{p,rad}} \int_{\mathbb{R}} (v_r-v_{p,rad}) v_\phi^\gamma \delta\left( v_r - \Tilde{v} \right) \,dv_r\,dv_\phi
    \\ &= \frac{1}{2} \int_{\sqrt{\frac{(1 - e)\mu}{r}}}^{\sqrt{\frac{(1 + e)\mu}{r}}} (\Tilde{v}-v_{p,rad}) v_\phi^\gamma H(-\Tilde{v} + v_{p,rad}) \,dv_\phi,
    \label{eq:j_rad_minus_post}
\end{split}\end{equation}
where 
\begin{equation}
    \tilde{v} = \frac{\sqrt{(e^2-1) \mu^2 + 2 \mu v_\phi^2 r - v_\phi^4 r^2}}{v_\phi r},
\end{equation}
and altogether:
\begin{equation}
    j_{tot,rad} = C \delta_0(z) r^\gamma \left( j_{rad}^{+,pre} + j_{rad}^{+,post} - j_{rad}^{-,pre} - j_{rad}^{-,post} \right). \label{eq:j_tot_rad}
\end{equation}

We note that even and odd terms are straightforward to join as the only difference is the complementary Heaviside, but in the present shape it is easy to account for different effective areas from front and from the back of the probe in this function, since front and back terms are separated. Eqs.~\ref{eq:j_rad_plus_pre} -- \ref{eq:j_rad_minus_post} are straight-forward to evaluate numerically, for example with Monte Carlo integration, drawing values of $v_{phi}$ between the boundaries (Eq.~\ref{eq:derivation_integration_boundaries}).

\paragraph{Azimuthal flux}
\begin{equation}\begin{split}
    j_{tot,azim} &= \delta(z) \int_{\mathbb{R}} \int_{v_{p,azim}}^\infty (v_\phi-v_{p,azim}) f(r,v_r,v_\phi)  \,dv_r\,dv_\phi
    \\ &\qquad - \delta(z) \int_{\mathbb{R}} \int_{-\infty}^{v_{p,azim}} (v_\phi-v_{p,azim}) f(r,v_r,v_\phi)  \,dv_r\,dv_\phi,
    \\ &= \delta(z) C r^\gamma \int_{\mathbb{R}} \int_{v_{p,azim}}^\infty (v_\phi-v_{p,azim}) v_\phi^\gamma \delta\left( v_r \pm \Tilde{v} \right)  \,dv_r\,dv_\phi
    \\ &\qquad - \delta(z) C r^\gamma \int_{\mathbb{R}} \int_{-\infty}^{v_{p,azim}} (v_\phi-v_{p,azim}) v_\phi^\gamma \delta\left( v_r \pm \Tilde{v} \right)  \,dv_r\,dv_\phi
\end{split}\end{equation}
The expression contains two terms: $j_{tot,azim} = \delta(z) C r^\gamma ( j_{azim}^+ - j_{azim}^- )$, which have the boundaries $v_\phi > v_{p,azim}$ and $v_{p,azim} > v_\phi$ respectively. Since we assume prograde dust only ($v_\phi>0$), and pre-perihelion and post-perihelion have the same effect on the azimuthal flux, there is no further multiplication of terms, as in the case of radial flux. 
\begin{equation}\begin{split}
    j_{azim}^+ &= \int_{\mathbb{R}} \int_{v_{p,azim}}^\infty (v_\phi-v_{p,azim}) v_\phi^\gamma \delta\left( v_r \pm \Tilde{v} \right)  \,dv_r\,dv_\phi
    \\ &= \int_{v_{p,azim}}^\infty (v_\phi-v_{p,azim}) v_\phi^\gamma \,dv_\phi
    \\ & = \int_{\max{\left[\sqrt{\frac{(1 - e)\mu}{r}},v_{p,azim}\right]}}^{\max{\left[\sqrt{\frac{(1 + e)\mu}{r}},v_{p,azim}\right]}} (v_\phi - v_{p,azim}) v_\phi^{\gamma} \,dv_\phi
    \\ &= \left[ \frac{v_\phi^{\gamma+2}}{\gamma+2} - \frac{v_\phi^{\gamma+1}v_{p,azim}}{\gamma+1} \right]_{\max{\left[\sqrt{\frac{(1 - e)\mu}{r}},v_{p,azim}\right]}}^{\max{\left[\sqrt{\frac{(1 + e)\mu}{r}},v_{p,azim}\right]}},
    \label{eq:j_azim_plus}
\end{split}\end{equation}
\begin{equation}\begin{split}
    j_{azim}^- &= \int_{\mathbb{R}} \int_{-\infty}^{v_{p,azim}} (v_\phi-v_{p,azim}) v_\phi^\gamma \delta\left( v_r \pm \Tilde{v} \right)  \,dv_r\,dv_\phi
    \\ &= \int_{-\infty}^{v_{p,azim}} (v_\phi-v_{p,azim}) v_\phi^\gamma \,dv_\phi
    \\ & = \int_{\min{\left[\sqrt{\frac{(1 - e)\mu}{r}},v_{p,azim}\right]}}^{\min{\left[\sqrt{\frac{(1 + e)\mu}{r}},v_{p,azim}\right]}} (v_\phi - v_{p,azim}) v_\phi^{\gamma} \,dv_\phi
    \\ &= \left[ \frac{v_\phi^{\gamma+2}}{\gamma+2} - \frac{v_\phi^{\gamma+1}v_{p,azim}}{\gamma+1} \right]_{\min{\left[\sqrt{\frac{(1 - e)\mu}{r}},v_{p,azim}\right]}}^{\min{\left[\sqrt{\frac{(1 + e)\mu}{r}},v_{p,azim}\right]}},
    \label{eq:j_azim_minus}
\end{split}\end{equation}
where we very liberally ignred $\delta_0(v_r \pm \dots)$, but since we integrate in $v_r$ over $\mathbb{R}$, it doesn't matter where exactly this mass is accounted for. Altogether we have: 
\begin{equation}
    j_{tot,azim} = \delta(z) C r^\gamma ( j_{azim}^+ - j_{azim}^- ), \label{eq:j_tot_azim}
\end{equation}
which is easy and straightforward to evaluate. Finally,
\begin{equation}
    j_{tot} = j_{tot,azim} + j_{tot,rad}.
\end{equation}

\subsection{Normalization} \label{ch:a:normalization}

In order to normalize the flux properly to a known value of density at $1\,\si{AU}$ in the unit of $\si{m^{-3}}$, we need to evaluate the number density at $r_0$, which might conveniently be $1\,\si{AU}$. If we don't do that, then Eqs.~\ref{eq:j_rad_plus_pre} - \ref{eq:j_rad_minus_post}, \ref{eq:j_azim_plus}, \ref{eq:j_azim_minus} all vary as $\propto 2\sqrt{e}$ for low $e$. 
Let's evaluate $n(r=r_0)$ for the parameter $e$. Analogically to Eqs.~\ref{eq:j_tot_rad} and $\ref{eq:j_tot_azim}$:
\begin{equation}\begin{split}
    n &= \delta(z) \int_{\mathbb{R}} \int_{\mathbb{R}} f(r,v_r,v_\phi)  \,dv_r\,dv_\phi,
    \\ &= \delta(z) C r_0^\gamma \int_{\mathbb{R}} \int_{\mathbb{R}} v_\phi^\gamma \delta\left( v_r \pm \Tilde{v} \right)  \,dv_r\,dv_\phi
    \\ &= \delta(z) C r_0^\gamma  \int_{\sqrt{\frac{(1 - e)\mu}{r_0}}}^{\sqrt{\frac{(1 + e)\mu}{r_0}}} v_\phi^\gamma \,dv_\phi
    \\ &= \delta(z) C r_0^\gamma  \left[ \frac{v_\phi^{\gamma+1}}{\gamma+1} \right]_{\sqrt{\frac{(1 - e)\mu}{r_0}}}^{\sqrt{\frac{(1 + e)\mu}{r_0}}}
    \\ &= \delta(z) C r_0^\gamma \frac{1}{\gamma+1} \left[ v_\phi^{\gamma+1} \right]_{\sqrt{\frac{(1 - e)\mu}{r_0}}}^{\sqrt{\frac{(1 + e)\mu}{r_0}}}
    \\ &= \delta(z) C r_0^\gamma \frac{1}{\gamma+1} \left( \left(\frac{(1 + e)\mu}{r_0}\right)^{\frac{\gamma+1}{2}} - \left(\frac{(1 - e)\mu}{r_0}\right)^{\frac{\gamma+1}{2}} \right)
    \\ &= \delta(z) C r_0^\gamma \frac{1}{\gamma+1} \left( {\frac{\mu}{r_0}} \right)^{\frac{\gamma+1}{2}} \left( (1 + e)^{\frac{\gamma+1}{2}} - (1 - e)^{\frac{\gamma+1}{2}} \right)
\end{split}\end{equation}
Thus, we get that:
\begin{equation}
    \delta_0(z) C = \frac{n}{r_0^\gamma} \left( {\frac{r_0}{\mu}} \right)^{\frac{\gamma+1}{2}}  \frac{ (\gamma+1)}{\left( (1 + e)^{\frac{\gamma+1}{2}} - (1 - e)^{\frac{\gamma+1}{2}} \right)}, 
\end{equation}
where $[n]=m^{-3}$ at the distance of $r_0$.

\subsection{Number density assumption} \label{ch:a:number_density_validity}

\begin{figure}[h]
 	\centering
 	\includegraphics[width=9cm]{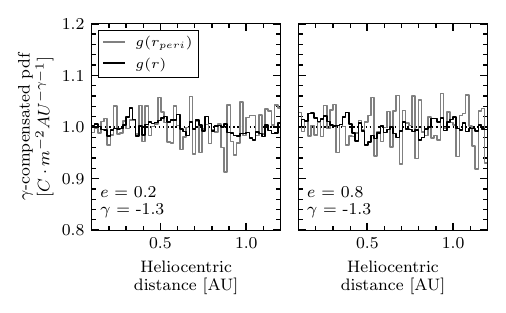}
 	\caption{A sample of $r_{peri}$ is drawn according to $n \propto r_{peri}^\gamma$ and transformed using $g(r|r_{peri})$. The distribution shows $n \propto r^\gamma$.}
 	\label{fig:number_density_scaling}
\end{figure}

To derive the fluxes following the assumption of number density scaling as $n \propto r^\gamma$, we conveniently assumed the equivalence 
\begin{equation}
    n \propto r^\gamma \Leftrightarrow n \propto r_{peri}^\gamma.
\end{equation}
Here we demonstrate the validity of this assumption. Assume a dust grain is in orbit with the perihelion $r_{peri}$ and aphelion 
\begin{equation}
    r_{aph} = \frac{1+e}{1-e} r_{peri},
\end{equation}
where $e$ is the eccentricity. The grain therefore spends time at the heliocentric distance $r$:
\begin{equation}
    r_{peri} < r < r_{aph},
\end{equation}
and the time spent, and therefore the probability $g(r)$ the grain will be (in random time) found at $r$ it spends at $r$ is proportional to the inverse of the radial speed  $|v_r|$:
\begin{equation}
    g(r|r_{peri}) \propto |v_r|^{-1} (r) = \left| \frac{dr}{dt}(r) \right|^{-1} = \left( v^2 - v_{\phi}^2 \right)^{-\frac{1}{2}},
\end{equation}
where $v$ and $v_{\phi}$ are the total and azimuthal speeds of the grain. Then we have from vis-viva equation:
\begin{equation}
    v^2 = \mu \left( \frac{2}{r} - \frac{1}{a} \right) = \mu \left( \frac{2}{r} - \frac{1-e}{r_{peri}} \right),
\end{equation}
and $v_{\phi}$ is obtained using momentum conservation as
\begin{equation}
    v_{\phi} = \frac{v_{peri} r_{peri}}{r},
\end{equation}
where the perihelion speed $v_{peri}$ is also obtained from vis-viva as
\begin{equation}
    v_{peri} = \sqrt{\mu \frac{1+e}{r_{peri}}},
\end{equation}
which altogether gives:
\begin{equation}\begin{split}
   g(r|r_{peri}) \propto |v_r| (r) &= \left(\mu \left( \frac{2}{r} - \frac{1-e}{r_{peri}} \right) - \left( \frac{\sqrt{\mu \frac{1+e}{r_{peri}}} r_{peri}}{r} \right)^2\right)^{-\frac{1}{2}} \\ 
    &= \left(\mu \left( \frac{2}{r} - \frac{1-e}{r_{peri}} -  \frac{(1+e) r_{peri}}{r^2} \right) \right)^{-\frac{1}{2}} \\
    &= \left(\frac{2\mu}{r} \left( 1 - \left( \frac{(1-e)}{2}\frac{r}{r_{peri}} + \frac{(1+e)}{2}\frac{r_{peri}}{r} \right) \right) \right)^{-\frac{1}{2}}. 
\end{split}\end{equation}
Obtaining $r$ from $r_{peri}$ is a random process governed by the probability density function $g(r|r_{peri})$. We can therefore draw a sample of $r_{peri}$ according to $g(r_{peri})$ and transform $r_{peri}$ to $r$ using the density $g(r|r_{peri})$ derived here. Fig.~\ref{fig:number_density_scaling} shows the $\gamma$ compensated probability density function of $g(r_{peri})$ before the transformation and $g(r)$ after the transformation. A compensated density plot shows a constant, if the slope of the density is the value, for which we compensate, that is $\gamma$ in our case. As is observed, the dependence $n\propto r^\gamma$ is really retained after the transformation, if $n\propto r_{peri}^\gamma$ is assumed.

\subsection{Generalization} \label{ch:a:generalization}

\paragraph{Already included parameters}

The integrals for flux $j_{tot,rad}$ and $j_{tot,azim}$ as derived and expressed in Sec.~\ref{ch:a:integration} already allow for evaluation of the flux along a trajectory of a spacecraft, given the eccentricity of the dust's orbits $e$, radiation pressure to gravity ratio $\beta$, distance-scaling parameter $\gamma$ and cuboid-approximated spacecraft areas, since the flux is evaluated for each of the relevant spacecraft sides independently. 

\paragraph{Retrograde grains}

The simplest addition is the fraction of retrograde grains in the dust cloud. Until, now all the dust grains were considered prograde, but the fraction of retrograde dust is taken into account by weighted summing of the flux encountered along the true spacecraft trajectory, with the flux encountered by mirrored (retrograde) spacecraft trajectory. 

\paragraph{Higher exponent of the relative speed}

If the detection is effective regardless of the impact speed, the flux is proportional to the relative speed between the spacecraft and the dust cloud. If the flux is assumed proportional to an exponent $\epsilon$ of velocity, which is higher than unity, such as because of the detection threshold size variation with the impact speed, this is taken into account by evaluating the higher moment of $(v_r - v_{p,rad})$ and $(v_\phi - v_{p,azim})$ in the terms of Eqs.~\ref{eq:j_tot_rad} and \ref{eq:j_tot_azim} respectively, which is straight forward. We note that the normalization needs to be adjusted in this case as well, as a scale relative velocity $v_{0}$ has to be introduced, at which the number density is measured.

\paragraph{Inclination}

A single inclination angle $\theta$ different from zero can be introduced, under the assumption that all the grains share the same inclination value, albeit in different (non-parallel) orbital planes, that is with random ascending nodes. Under this assumption, a spacecraft in the plane of ecliptics will only encounter dust grains of this given inclination $\theta$. As discussed in Sec.~\ref{ch:dust_solo}, it is reasonable to approximate PSP by a cylinder. This makes no difference compared to a cuboid approximation, until $\theta\neq 0$ is examined. 

We note that an arbitrary inclination of the orbital plane of each of the grains does not play a role in $j_{tot,rad}$. For the azimuthal component, we need to evaluate the moment $|\vec{v}_\phi - \vec{v}_{p,azim}|$ over the $f$, as a function of inclination $\theta$. We assume the cylindrical symmetry:
\begin{equation}
    v_{cyl}(v_\phi) \equiv |\vec{v}_\phi - \vec{v}_{p,azim}| = \sqrt{v_{p,azim}^2 \sin^2\theta + \left(v_\phi - v_{p,azim}\cos\theta \right)^2},
\end{equation}
which we then need to integrate to
\begin{equation}
    j_{tot,azim} = \delta(z) C r^\gamma ( j_{azim}^+ - j_{azim}^- ),
\end{equation}
where
\begin{equation}
    j_{azim}^+ = \int_{\sqrt{\frac{(1 - e)\mu}{r}}}^{\sqrt{\frac{(1 + e)\mu}{r}}} v_{cyl}(v_\phi) v_\phi^\gamma H\left( v_{cyl}(v_\phi) \right)\,dv_\phi,
\end{equation}
\begin{equation}
    j_{azim}^-= \int_{\sqrt{\frac{(1 - e)\mu}{r}}}^{\sqrt{\frac{(1 + e)\mu}{r}}} v_{cyl}(v_\phi) v_\phi^\gamma 
    H\left( - v_{cyl}(v_\phi) \right)\,dv_\phi.
\end{equation}
Since by definition $| \cdot | >0$:
\begin{equation}
    j_{azim}^+ - j_{azim}^- = \int_{\sqrt{\frac{(1 - e)\mu}{r}}}^{\sqrt{\frac{(1 + e)\mu}{r}}} v_{cyl}(v_\phi) v_\phi^\gamma \,dv_\phi,
\end{equation}
which we evaluate easily, for example with Monte Carlo integration, drawing values of $v_{phi}$ between the integration boundaries (Eq.~\ref{eq:derivation_integration_boundaries}).

We note that both non-zero eccentricity and non-zero inclination make the assumption of non-interacting grains problematic, but the collisional evolution of the dust cloud is beyond the scope of this work, and taken care of in reality by the micrometer dust cloud being constantly replenished by the product of collision of bigger grains.

\section{Model trajectories of PSP} \label{app:ephemeris}

Every PSP's solar encounter is different from the previous one, even within the same orbital group. This is solely because of the motion of the Sun, which in the first approximation orbits around the common barycenter of the Sun --- Jupiter system, which lies outside of the solar photosphere. PSP orbits the Sun on an orbit with perihelion distance sufficiently close the the Sun, so that this effect plays a role when the distance from the Sun is critical. This is however not as consequential as to change the results presented in this work. To get rid of the effect, we study fictitious, simplified solar encounters, which are assumed to lie in the ecliptic plane ($z=0$) and which represent the actual ones well. The parameters of the encounters which we use for the present study are listen in Tab.~\ref{tab:encounter_paramters}.

\begin{table}
\caption{The representative orbital parameters for PSP's encounter groups.}
\centering
\label{tab:encounter_paramters}
\begin{tabular}{cccc}
\hline\hline
\multicolumn{1}{p{1.5cm}}{ \centering Encounter \\ group} & \multicolumn{1}{p{2cm}}{ \centering Perihelion \\ distance [Gm]} & \multicolumn{1}{p{2cm}}{ \centering Perihelion \\ distance [AU]} & \multicolumn{1}{p{1.8cm}}{ \centering Perihelion \\ speed [km/s]} \\
\hline
$1$ & $24.8$ & $0.166$ & $95 $ \\
$2$ & $19.4$ & $0.130$ & $109$ \\
$3$ & $14.2$ & $0.095$ & $127$ \\
$4$ & $11.1$ & $0.074$ & $147$ \\
$5$ & $9.2 $ & $0.061$ & $163$ \\ 
\hline
\end{tabular}
\end{table}

\section{The flux slope scaling with orbital parameters} \label{app:individual_parameters_slope}

\begin{figure}[h]
 	\centering
 	\includegraphics[width=9cm]{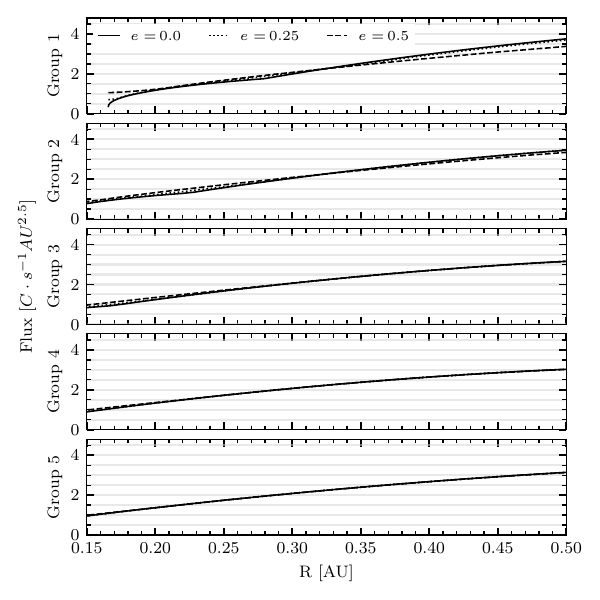}
 	\caption{The base model-predicted flux is shown in the solid line. In addition, the influence of eccentricity on the slope is demonstrated.}
 	\label{fig:compensated_e}
\end{figure}

\begin{figure}[h]
 	\centering
 	\includegraphics[width=9cm]{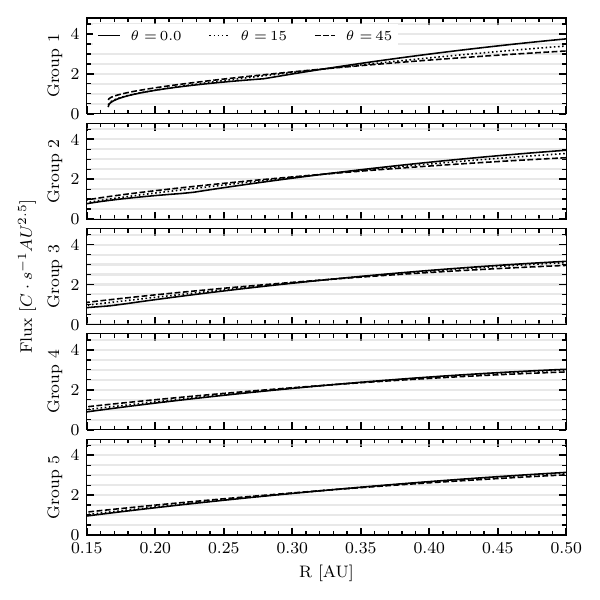}
 	\caption{The base model-predicted flux is shown in the solid line. In addition, the influence of inclination on the slope is demonstrated.}
 	\label{fig:compensated_theta}
\end{figure}

\begin{figure}[h]
 	\centering
 	\includegraphics[width=9cm]{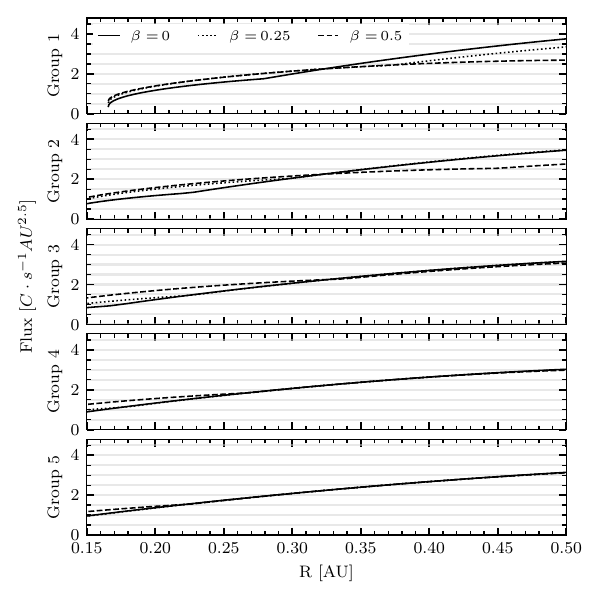}
 	\caption{The base model-predicted flux is shown in the solid line. In addition, the influence of $\beta$ value on the slope is demonstrated.}
 	\label{fig:compensated_beta}
\end{figure}

\begin{figure}[h]
 	\centering
 	\includegraphics[width=9cm]{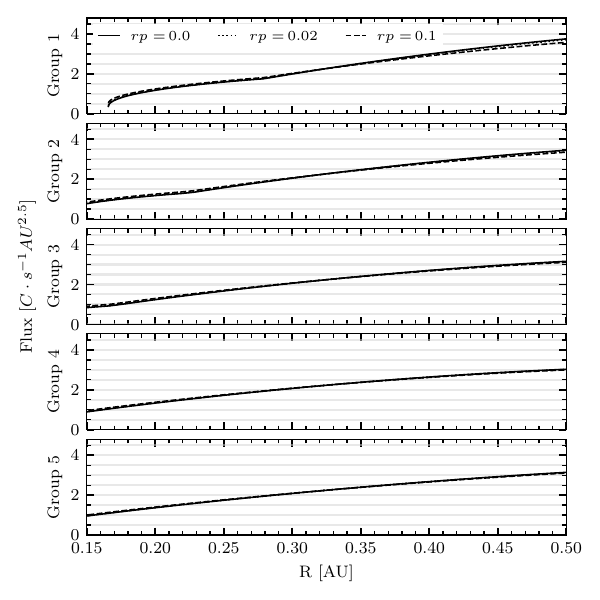}
 	\caption{The base model-predicted flux is shown in the solid line. In addition, the influence of retrograde fraction on the slope is demonstrated.}
 	\label{fig:compensated_rp}
\end{figure}

In Sec.~\ref{ch:scaling}, we claimed that the parameters: $e; \theta; \beta; rp$ all act to make the slope of flux steeper, when their values differ from the base case, but in the case of the later orbits, they tend to lead to very little change. Their influence is independently shown in Figs.~\ref{fig:compensated_e} -- \ref{fig:compensated_rp}.

\section{The near-perihelia flux dependence on other parameters} \label{app:individual_parameters_perihelia}

\begin{figure}[h]
 	\centering
 	\includegraphics[width=9cm]{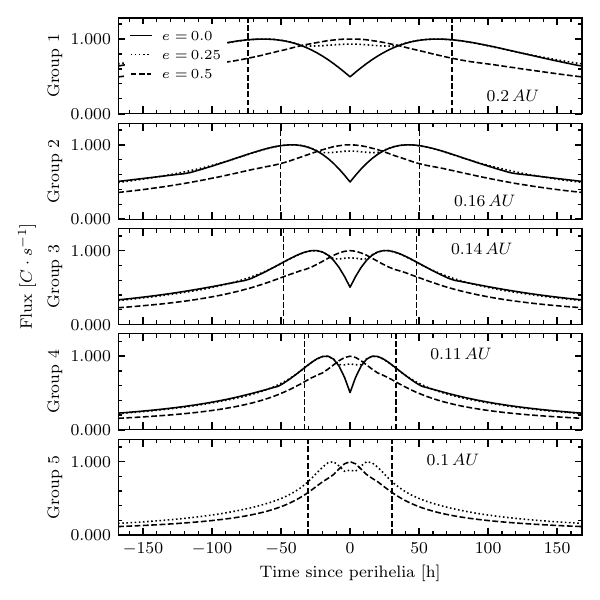}
 	\caption{The base model-predicted flux is shown in the solid line. In addition, the influence of eccentricity on the slope is demonstrated.}
 	\label{fig:perihelia_e}
\end{figure}

\begin{figure}[h]
 	\centering
 	\includegraphics[width=9cm]{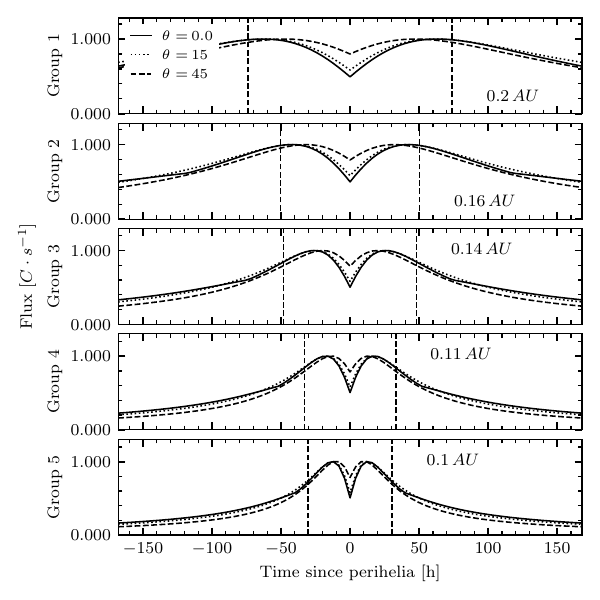}
 	\caption{The base model-predicted flux is shown in the solid line. In addition, the influence of inclination is shown.}
 	\label{fig:perihelia_theta}
\end{figure}

\begin{figure}[h]
 	\centering
 	\includegraphics[width=9cm]{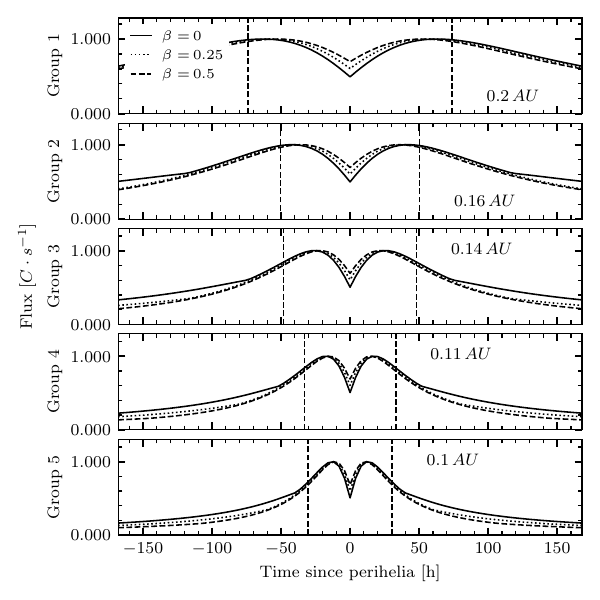}
 	\caption{The base model-predicted flux is shown in the solid line. In addition, the influence of $\beta$ value is shown.}
 	\label{fig:perihelia_beta}
\end{figure}

\begin{figure}[h]
 	\centering
 	\includegraphics[width=9cm]{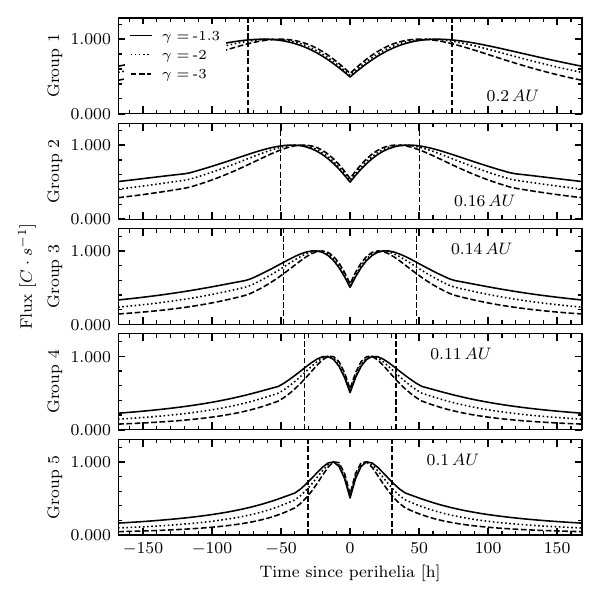}
 	\caption{The base model-predicted flux is shown in the solid line. In addition, the influence of the spatial density scaling with heliocentric distance $\gamma$ is shown.}
 	\label{fig:perihelia_gamma}
\end{figure}

\begin{figure}[h]
 	\centering
 	\includegraphics[width=9cm]{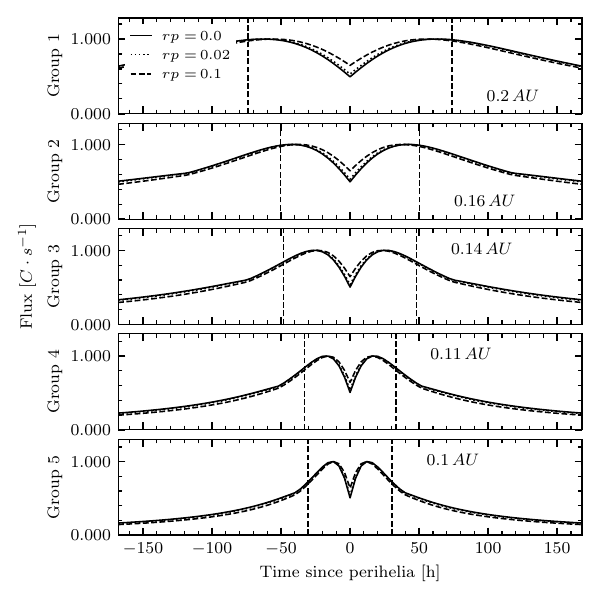}
 	\caption{The base model-predicted flux is shown in the solid line. In addition, the influence of retrograde fraction is shown.}
 	\label{fig:perihelia_rp}
\end{figure}

In Sec.~\ref{ch:perihelia}, we claimed that the parameters: $e; \theta; \beta; \gamma, rp$ do not change the location of the flux maxima appreciably. They however make the magnitude of the near-perihelia dip smaller, especially the parameter $e$ does. Their influence is independently shown in Figs.~\ref{fig:perihelia_e} -- \ref{fig:perihelia_rp}.

\end{appendix}

\end{document}